\documentclass[prd,preprint,tightenlines,floatfix,showpacs,preprintnumbers,nofootinbib,eqsecnum,superscriptaddress]{revtex4}

 \usepackage[dvips,final]{graphicx}
  \usepackage{amssymb}
   \usepackage{amsmath}
    \usepackage{amsfonts}
     \usepackage{epsfig}
      \usepackage{bm}% bold math

\usepackage[section]{placeins}

\usepackage{multirow}
\usepackage{ctable}
\usepackage{booktabs}
\usepackage{array}
\usepackage{tabularx}
\usepackage{xcolor}
\usepackage{pstricks}
\newcommand{\bp}{\mbox{$\vec{p}_T$}}
\newcommand{\bq}{\mbox{$\vec{q}_T$}}

\newcommand{\qperp}{\mbox{$\vec{q}_\perp$}}
\newcommand{\eperp}{\mbox{$\vec{e}_\perp$}}

\newcommand{\aem}{{\alpha_{\mathrm{em}}}}

\def\lsim{\mathrel{\rlap{\lower4pt\hbox{\hskip1pt$\sim$}}
    \raise1pt\hbox{$<$}}}         %less than or approx. symbol
\def\gsim{\mathrel{\rlap{\lower4pt\hbox{\hskip1pt$\sim$}}
    \raise1pt\hbox{$>$}}}         %greater than or approx. symbol

\begin{document}

\vfill
\title{Production of $W^+ W^-$ pairs via $\gamma^*\gamma^* \to W^+ W^-$
  subprocess \\
with photon transverse momenta}

\author{Marta {\L}uszczak}
\email{luszczak@univ.rzeszow.pl} \affiliation{
Faculty of Mathematics and Natural Sciences,
University of Rzesz\'ow, ul. Pigonia 1, PL-35-310 Rzesz\'ow, Poland}

\author{Wolfgang Sch\"afer}
\email{Wolfgang.Schafer@ifj.edu.pl} \affiliation{Institute of Nuclear Physics Polish Academy of Sciences, ul. Radzikowskiego 152, PL-31-342 Cracow, Poland}

\author{Antoni Szczurek}
\email{antoni.szczurek@ifj.edu.pl} \affiliation{Institute of Nuclear Physics Polish Academy of Sciences, ul. Radzikowskiego 152, PL-31-342 Cracow, Poland}

\date{\today}

\begin{abstract}
We discuss production of $W^+ W^-$ pairs in proton-proton collisions 
induced by two-photon fusion including, for a first time, 
transverse momenta of incoming photons.
The unintegrated inelastic fluxes (related to proton dissociation) of photons 
are calculated based on modern parametrizations of deep inelastic 
structure functions in a broad range of their arguments ($x$ and $Q^2$).
In our approach we can get separate contributions of different
$W$ helicities states. Several one- and two-dimensional differential 
distributions are shown and discussed. The present results are compared
to the results of previous calculations within collinear factorization
approach. Similar results are found except of some observables
such as e.g. transverse momentum of the pair of $W^+$ and $W^-$.
We find large contributions to the cross section from the region of 
large photon virtualities. We show decomposition of the total cross section as well as 
invariant mass distribution into the polarisation states of both W bosons.
The role of the longitudinal $F_L$ structure function is quantified.
Its inclusion leads to a 4-5 \% decrease of the cross section, almost independent of $M_{WW}$.
\end{abstract}


%----------------------------------------------------------------------

\maketitle

%---------------------------
\section{Introduction}
%---------------------------

Recently the partonic processes initiated by one or two photons in hadronic
collisions at the Large Hadron Collider (LHC) are
becoming an active field of research.  The corresponding theoretical 
approach requires the calculation of photon fluxes in the proton-proton
collision.
The majority of practical approaches focused on a collinear factorization
approach where the momentum of the colliding photon is collinear to 
the parent proton momentum.
For a comprehensive review on photon-photon fusion reactions, see
\cite{Budnev:1974de}.
Recently, for the conditions of LHC, photon-photon fusion was discussed 
in the context of lepton pairs \cite{daSilveira:2014jla,Luszczak:2015aoa}, 
$W^+ W^-$ \cite{LSR2015} or possible signals beyond the Standard Model, 
such as the production of charged Higgs bosons $H^+ H^-$ \cite{LS2015}.
In Ref.\cite{LSR2015} it was shown that photon-photon partonic processes
are 
%extremely 
important for large invariant masses of $W^+ W^-$ pairs.

Several groups that provide the high-energy community with
parton distribution functions included photons as partons in 
the proton \cite{Martin:2004dh,Ball:2013hta,Schmidt:2015zda,Giuli:2017oii},
solving the corresponding coupled DGLAP evolution equations.

This strategy differs from the one adopted in Ref. \cite{daSilveira:2014jla,Luszczak:2015aoa}
(see also Ref.\cite{Ginzburg:1998vb}),
where following Ref. \cite{Budnev:1974de}, the photon fluxes had been calculated
in a data-driven way using their relation to the well-measured proton
structure functions. Subsequently, such a data-driven approach was taken up in
Refs. \cite{Manohar:2016nzj,Manohar:2017eqh}.

The transverse momenta of photons were included so far
only for $\gamma \gamma \to e^+ e^-$ or $\gamma \gamma \to \mu^+ \mu^-$
subprocesses \cite{daSilveira:2014jla,Luszczak:2015aoa}. 
There we identified corners of phase space where transverse momenta 
of photons (or their virtualities) are large. 

In the present paper we extend our studies to the production of $W^+ W^-$ pairs.
We expect that here the virtualities of photons may be much larger
than for $l^+ l^-$ production.

\begin{figure*}
\begin{center}
\includegraphics[width=5cm,height=4cm]{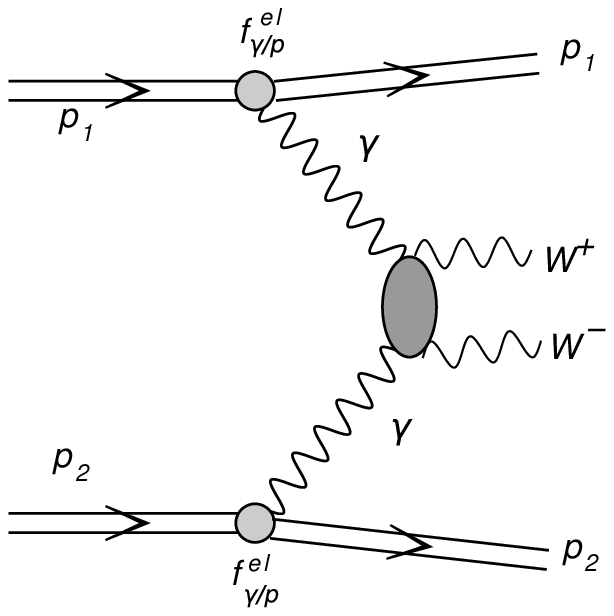}
\includegraphics[width=5cm,height=4cm]{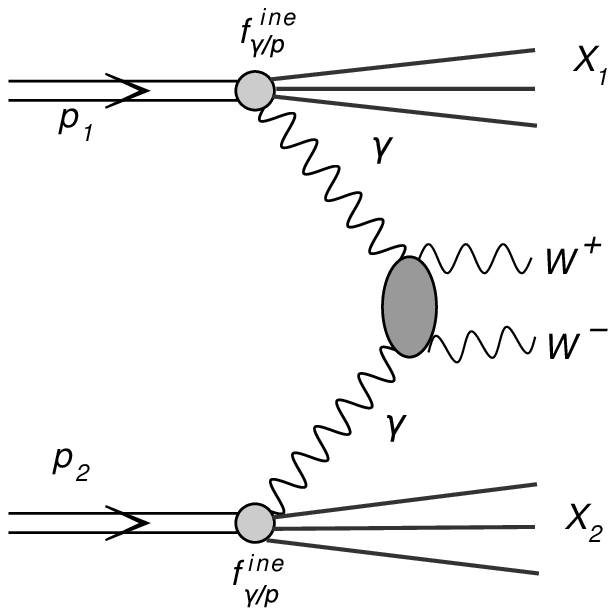} \\
\includegraphics[width=5cm,height=4cm]{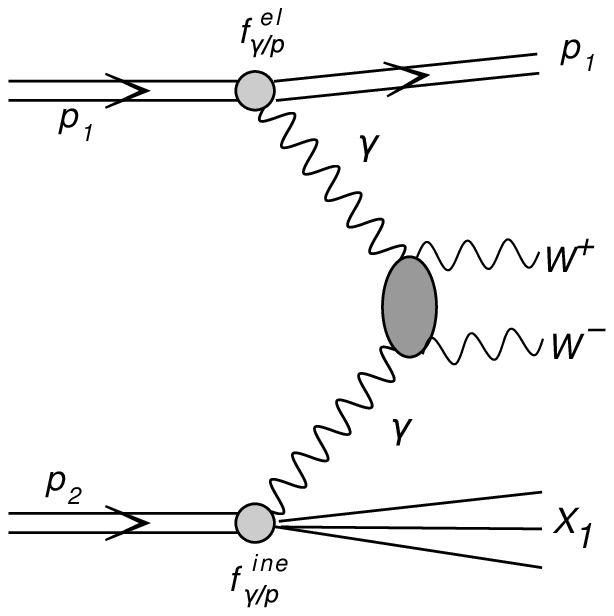}
\includegraphics[width=5cm,height=4cm]{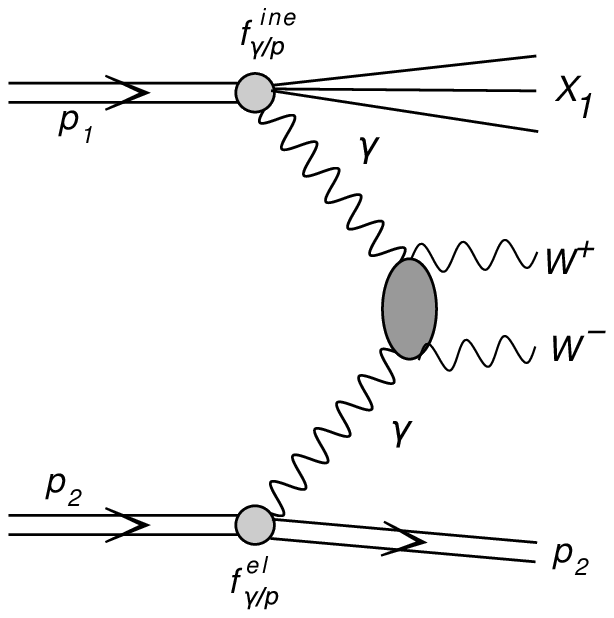}
\caption{Diagrams representing different categories of photon-photon 
induced mechanisms for production of $W^+ W^-$ pairs.
}
\label{fig:new_diagrams}
\end{center}
\end{figure*}

Particularly interesting is the region
of large invariant masses of the $W^+ W^-$ system where 
the diphoton mechanism becomes
one of the most important contributions for $W^+ W^-$ pair production.
We shall compare the calculation within the $k_T$-factorization approach
with those obtained previously in the collinear approximation.
We shall discuss all types of processes as shown in 
Fig.\ref{fig:new_diagrams}.

The $\gamma \gamma \to W^+ W^-$ subprocess is interesting also
in the context of searches of effects beyond Standard Model effects 
\cite{Chapon:2009hh,Pierzchala:2008xc}, such as anomalous 
quartic gauge-boson couplings.
First experimental studies on anomalous $\gamma\gamma WW$ couplings 
were already presented recently both by the CMS and ATLAS collaborations
\cite{Khachatryan:2016mud,Aaboud:2016dkv}.
We expect that our present estimate within the Standard Model 
will be therefore a useful reference point in searches beyond Standard Model.
We shall also present a separate contribution for longitudinal $W$ boson
which is interesting in the contex of $WW$ final state interactions
and/or searches for possible resonances, see for example  \cite{Kilian:2015opv,Delgado:2016rtd,Szleper:2014xxa}.

%=================================================================================
%===============================================================

%---------------------------------------------------------------------------
\section{Accounting for transverse momenta of photons}
\label{sec:fluxes}
%---------------------------------------------------------------------------

In this section we will include the transverse momentum of photons, so that
the distributions of the transverse momentum of the $W^+ W^-$ pair and
the azimuthal angle between the $W$'s have a nontrivial behaviour already 
at the lowest order.
In \cite{daSilveira:2014jla,Luszczak:2015aoa} a $k_T$-factorization approach
for the $\gamma \gamma$-fusion reactions in the high-energy limit
of the $pp$-collision has been given. This approach has its domain of 
applicability
in the region of small momentum fractions $z \ll 1$ carried by photons.

In this case, the unintegrated photon distributions can be calculated from 
the proton structure function $F_2(x,Q^2)$ alone in a data-driven way.

A broader range of applicability has the generalized equivalent-photon 
approximation of \cite{Budnev:1974de}, in which a whole density matrix 
of photons appears.
In some instances, for example when the masses squared of produced particles
are much larger than the typical virtualities of photons, the density matrix
simplifies and only transverse polarizations in the center-of-mass of 
colliding photons are important.
We will adopt this approach for our numerical calculations of 
$W^+ W^-$ bosons below.

In both $k_T$-dependent approaches described above, the cross section
for $W^+W^-$ production can be written in the form 
\begin{eqnarray}
{d \sigma^{(i,j)} \over dy_1 dy_2 d^2\bp_1 d^2\bp_2} &&=  \int  {d^2 \bq_1 \over \pi \bq_1^2} {d^2 \bq_2 \over \pi \bq_2^2}  
{\cal{F}}^{(i)}_{\gamma^*/A}(x_1,\bq_1) \, {\cal{F}}^{(j)}_{\gamma^*/B}(x_2,\bq_2) 
{d \sigma^*(p_1,p_2;\bq_1,\bq_2) \over dy_1 dy_2 d^2\bp_1 d^2\bp_2} \, , \nonumber \\ 
\label{eq:kt-fact}
\end{eqnarray}
where the indices $i,j \in \{\rm{el}, \rm{in} \}$ denote elastic or inelastic final states.
The longitudinal momentum fractions of photons are obtained from 
the rapidities and transverse momenta of final state $W^+W^-$ as:
\begin{eqnarray}
x_1 &=& \sqrt{ {\bp_1^2 + m_W^{2} \over s}} e^{y_1} +  \sqrt{ {\bp_2^2 +
		m_W^2 \over s}} e^{y_2} 
\; , \nonumber \\
x_2 &=& \sqrt{ {\bp_1^2 + m_W^{2} \over s}} e^{-y_1} +  \sqrt{ {\bp_2^2 + m_W^2 \over s}} e^{-y_2} \, .
\end{eqnarray}

For photons which carry transverse polarization in the 
$\gamma \gamma$-cms frame, we write
the relevant ``off-shell'' cross section as:
%%%
\begin{eqnarray}
{d \sigma^*(p_1,p_2;\bq_1,\bq_2) \over dy_1 dy_2 d^2\bp_1 d^2\bp_2} = {1 \over 16 \pi^2 (x_1 x_2 s)^2} 
\sum_{\lambda_{W^+} \lambda_{W^-}} |M(\lambda_{W^+}, \lambda_{W^-})|^2 \, \delta^{(2)}(\bp_1 + \bp_2 - \bq_1 - \bq_2) \, ,
\nonumber \\
\end{eqnarray}
where the matrix element $M$ in terms of transverse momenta of incoming 
photons is given by
\begin{eqnarray}
M(\lambda_{W^+} \lambda_{W^-} )&=& 
{ 1 \over |\qperp_1||\qperp_2|} 
\sum_{\lambda_1 \lambda_2} (\eperp(\lambda_1) \cdot \qperp_1) 
(\eperp^*(\lambda_2) \cdot \qperp_2) 
{\cal{M}}(\lambda_1, \lambda_2; \lambda_{W^+}, \lambda_{W^-}) \nonumber \\
&=& { 1 \over |\qperp_1||\qperp_2|} \sum_{\lambda_1 \lambda_2} q_{\perp 1}^i q_{\perp 2}^j  \, e_i(\lambda_1) e_j^*(\lambda_2) 
\cdot {\cal{M}}(\lambda_1, \lambda_2; \lambda_{W^+}, \lambda_{W^-}),
\end{eqnarray}
with $\eperp(\lambda) = {-i}(\lambda \vec{e}_x + i \vec{e}_y)$. 
The helicity matrix elements 
${\cal{M}}(\lambda_1, \lambda_2; \lambda_{W^+}, \lambda_{W^-})$
for the process 
$\gamma(\lambda_1) \gamma(\lambda_2) \to W^+(\lambda_{W^+}) W^-(\lambda_{W^-})$ are taken from
Ref. \cite{Nachtmann:2005en}, where one can also find explicit helicity 
states defined in the cm-frame of the $W^+W^-$ pair. 
It is useful to decompose the matrix element further, using the identity
\begin{eqnarray}
q_{\perp 1}^i q_{\perp 2}^j &=& {1 \over2} \delta_{ij} (\qperp_1 \cdot \qperp_2) + 
{ 1 \over 2} \Big( q_{\perp 1}^i q_{\perp 2}^j + q_{\perp 1}^j q_{\perp 2}^i - \delta_{ij} (\qperp_1 \cdot \qperp_2) \Big)
+  { 1 \over 2} \Big( q_{\perp 1}^i q_{\perp 2}^j - q_{\perp 1}^j q_{\perp 2}^i \Big) \, \nonumber \\
&=& { 1 \over 2} \delta_{ij}   (\qperp_1 \cdot \qperp_2) + {1 \over 2} t^{kl}_{ij} q_{\perp 1}^k q_{\perp 2}^l 
+ { 1 \over 2} \epsilon_{ij} [\qperp_1, \qperp_2] \, .
\end{eqnarray}
Here the antisymmetric symbol is defined by 
$\epsilon_{xy} = - \epsilon_{yx} = 1$, and
\begin{eqnarray}
[\qperp_1, \qperp_2] \equiv q_{\perp 1}^x q_{\perp 2}^y -  q_{\perp 1}^y q_{\perp 2}^x \, , 
\end{eqnarray}
furthermore
\begin{eqnarray}
t_{ij}^{kl} = \delta^k_i \delta^l_j + \delta^k_j \delta^l_i - \delta^i_j \delta^k_l \, . 
\end{eqnarray}
We then obtain for the helicity-matrix element
\begin{eqnarray}
M(\lambda_{W^+} \lambda_{W^-} ) &=&  { 1 \over |\qperp_1||\qperp_2|} \Big\{
(\qperp_1 \cdot \qperp_2) \cdot \Big( {\cal{M}}(++;\lambda_{W^+} \lambda_{W^-}  ) + {\cal{M}}(--;\lambda_{W^+} \lambda_{W^-} )  \Big)
\nonumber \\
&-& i [\qperp_1, \qperp_2] \Big( {\cal{M}}(++;\lambda_{W^+} \lambda_{W^-}  ) - {\cal{M}}(--;\lambda_{W^+} \lambda_{W^-}) \Big)
\nonumber \\
&-& \Big( q_{\perp 1}^x q_{\perp 2}^x - q_{\perp 1}^y q_{\perp 2}^y \Big)
\Big( {\cal{M}}(+-;\lambda_{W^+} \lambda_{W^-}  ) + {\cal{M}}(-+;\lambda_{W^+} \lambda_{W^-}) \Big)
\nonumber \\
&-& i  \Big( q_{\perp 1}^x q_{\perp 2}^y + q_{\perp 1}^y q_{\perp 2}^x \Big) 
\Big( {\cal{M}}(+-;\lambda_{W^+} \lambda_{W^-}  ) - {\cal{M}}(-+;\lambda_{W^+} \lambda_{W^-}) \Big) \, .
\label{helicity_ME}
\end{eqnarray}
Together with these matrix elements, we use the photon fluxes from 
\cite{Budnev:1974de}.
We write the photon distribution differentially as
\begin{eqnarray}
dn^{\mathrm{in,el}} = {dz \over z} {d^2 \bq \over \pi \bq^2} \, {\cal{F}}^{\mathrm{in,el}}_{\gamma^* \leftarrow p} (z,\bq) \, .
\end{eqnarray}
The virtuality $Q^2$ of the photon carrying momentum fraction $z$ and transverse momentum $\vec{q}_T$ is
\begin{eqnarray}
Q^2 =  {\bq^2 + z (M_X^2 - m_p^2) + z^2 m_p^2 \over (1-z)} \, ,
\end{eqnarray}
where $M_X$ is the invariant mass of the proton remnant in the final state.
Then using 
%%%
%
%%
\begin{eqnarray}
 {dQ^2 \over Q^2} = {Q^2 - Q^2_{\rm min} \over Q^2}{d^2 \bq \over \pi \bq^2} , 
 \, {\rm and} \, \, { \bq^2 \over \bq^2 + z (M_X^2 - m_p^2) + z^2 m_p^2 }= {Q^2 - Q^2_{\rm min} \over Q^2} \, ,
\end{eqnarray}
%%%
we can write the fluxes from \cite{Budnev:1974de} as

\begin{eqnarray}
{\cal{F}}^{\mathrm{in}}_{\gamma^* \leftarrow p} (z,\bq) &=& {\alpha_{\rm em} \over \pi} 
\Big\{(1-z) \Big( {\bq^2 \over \bq^2 + z (M_X^2 - m_p^2) + z^2 m_p^2  }\Big)^2  
{F_2(x_{\rm Bj},Q^2) \over Q^2 + M_X^2 - m_p^2}  \nonumber \\
&+& {z^2 \over 4 x^2_{\rm Bj}}  
{\bq^2 \over \bq^2 + z (M_X^2 - m_p^2) + z^2 m_p^2  }
{2 x_{\rm Bj} F_1(x_{\rm Bj},Q^2) \over Q^2 + M_X^2 - m_p^2} \Big\} \, ,
\label{eq:flux_in}
\end{eqnarray}
and similarly for the elastic piece
\begin{eqnarray}
{\cal{F}}^{\mathrm{el}}_{\gamma^* \leftarrow p} (z,\bq) &=& {\aem \over \pi} 
\Big\{ (1-z)   \,\Big( {\bq^2 \over \bq^2 + z (M_X^2 - m_p^2) + z^2 m_p^2  }\Big)^2
{4 m_p^2 G_E^2(Q^2) + Q^2  G_M^2(Q^2) \over 4m_p^2 + Q^2} \nonumber \\
&+& {z^2 \over 4}  {\bq^2 \over \bq^2 + z (M_X^2 - m_p^2) + z^2 m_p^2  } G_M^2(Q^2) \Big\} \; .
\nonumber \\
\end{eqnarray}
These fluxes differ from the ones from 
Ref. \cite{daSilveira:2014jla,Luszczak:2015aoa}, which apply in the
high energy limit. The difference in these approaches is threefold: 
firstly, fluxes in Ref. \cite{daSilveira:2014jla,Luszczak:2015aoa}
also include a contribution from longitudinal polarizations of photons 
in the $\gamma^* \gamma^*$ cms, secondly
within the accuracy of the high-energy limit, the fluxes of 
\cite{daSilveira:2014jla,Luszczak:2015aoa}
depend on $F_2(x_{\rm Bj},Q^2)$ only, and thirdly 
these fluxes must be accompanied by the corresponding off-shell 
matrix element.
Notice that in (\ref{eq:flux_in}) instead of 
$F_2(x_{\rm Bj},Q^2),F_1(x_{\rm Bj},Q^2)$, 
one may use the pair $F_2(x_{\rm Bj},Q^2),F_L(x_{\rm Bj},Q^2)$, where
%%%
\begin{eqnarray}
F_L(x_{\rm Bj},Q^2) = \Big( 1 + {4 x_{\rm Bj}^2 m_p^2 \over Q^2} \Big) F_2(x_{\rm Bj},Q^2) - 2 x_{\rm Bj} F_1(x_{\rm Bj},Q^2)
\end{eqnarray}
%%%
is the longitudinal structure function of the proton.

%-----------------------------------------------------------
\section{Collinear-factorization approach}
%-----------------------------------------------------------

In some cases it can be sufficient to neglect the transverse momenta of 
partons.
Then photons are treated as collinear partons in a proton. 
Like other parton densities, the photon distribution $\gamma(z,\mu^2)$
is a function of the longitudinal momentum fraction $z$ carried by the photon
and the factorization scale $\mu^2$ of the hard process the photon 
participates in.

A number of parametrizations of the photon parton distributions have become
available recently 
\cite{Martin:2004dh,Ball:2013hta,Schmidt:2015zda,Giuli:2017oii,Manohar:2016nzj,Manohar:2017eqh}.
Most of them are based on including photons into the coupled DGLAP evolution
equations for quarks and gluons \cite{Martin:2004dh,Ball:2013hta,Schmidt:2015zda,Giuli:2017oii} and
attempt to extract the photon distributions from either global fits or fits to 
processes that are deemed to have a strong sensitivity to the photon 
distribution.
A different approach is taken in Ref.\cite{Manohar:2016nzj,Manohar:2017eqh}, where 
similarly to Ref.\cite{Luszczak:2015aoa} a data driven approach is taken.
An explicit coherent contribution is related to the electromagnetic 
form factors of a proton. A second contribution is related to the 
proton structure functions $F_2$ and $F_L$.

%----------------------------------------------------------------------------
%\subsection{From photon PDFs to cross section}
%----------------------------------------------------------------------------

In the collinear approach the photon-photon contribution
to inclusive cross section for $W^+ W^-$ production can be written as:
\begin{equation}
{d \sigma^{(i,j)} \over d y_1 d y_2 d^2 p_T} 
= {1 \over 16 \pi^2 (x_1 x_2 s)^2}\sum_{i,j} 
x_1 \gamma^{(i)}(x_1,\mu^2) 
x_2 \gamma^{(j)}(x_2,\mu^2)
\overline{ |{\cal M}_{\gamma \gamma \rightarrow W^+ W^-}|^2 }.
\label{collinear_factorization_formula}
\end{equation}
Here 
\begin{eqnarray}
x_1 &=&  \sqrt{p_T^2 + m_W^2 \over s} 
\Big( \exp(y_1) + \exp(y_2) \Big) \; , \nonumber \\
x_2 &=&  \sqrt{p_T^2 + m_W^2 \over s} 
\Big( \exp(-y_1) + \exp(-y_2) \Big) \; . 
\end{eqnarray}
Above indices $i$ and $j$ denote $i,j = \rm{el, in}$, i.e. they
correspond to elastic or inelastic components similarly as for 
the $k_T$-factorization discussed in section \ref{sec:fluxes} above.
The factorization scale is chosen as $\mu^2 = m_T^2 = p_T^2 + m_W^2$.

Calculations with collinear partons from eq. \ref{collinear_factorization_formula} have the drawback, 
that at the lowest order the produced two-body system is strictly in 
back-to-back kinematics.
Consequently the distribution in transverse momentum of the produced pair is
a delta-function. Similarly behaved the distribution 
of the azimuthal angle $\Delta \phi$ between the produced particles, which
is a delta function centered at $\Delta \phi = \pi$.

It should be made clear, however, that in Monte-Carlo simulations of 
the inclusive $W^+ W^-$-pair production, collinear cross sections, 
such as the one given by 
(\ref{collinear_factorization_formula}) can be embedded into events 
including e.g. initial state emissions from
parton showers, which will give a finite transverse momentum to the 
$W^+ W^-$-pair. The effect of highly virtual photons must then be accounted 
for by matching to higher order processes such as e.g. 
$q \gamma \to q W^+ W^-$ or $ q q \to q q  W^+ W^-$. 
The necessary rather sophisticated computational techniques are described e.g. 
in \cite{Alwall:2014hca}. We are not aware of calculations of the processes 
of interest here in this approach and prefer to stick to the more 
straightforward $k_T$-factorization described in the previous section.
Also, it should be noted that when we refer to the collinear
approximation in the remainder of the text, we always refer 
to calculations from Eq.(\ref{collinear_factorization_formula}).

%-------------------------------------
\section{Results}
%-------------------------------------

In this section we shall show our results for the $k_T$-factorization
approach. We shall concentrate first on the inelastic-inelastic contribution
(see Fig.\ref{fig:new_diagrams}). In the present paper we will not
include experimental cuts but rather consider full phase space
calculations.

We start from showing the cross sections using different parametrizations
of proton structure functions. 

Here we use the following options:

\begin{enumerate}
	\item the Abramowicz-Levy-Levin-Maor fit \cite{Abramowicz:1991xz,Abramowicz:1997ms} used previously also
	 in \cite{Luszczak:2015aoa}, abbreviated here ALLM.
	
	\item a newly constructed parametrization, which at $Q^2 > 9 \, \rm{GeV}^2$ uses an NNLO calculation 
	of $F_2$ and $F_L$ from NNLO MSTW 2008 partons \cite{Martin:2009iq}. It employs a useful code by the MSTW group \cite{Martin:2009iq}
	to calculate structure functions. At $Q^2 > 9 \, \rm{GeV}^2$ this fit uses the parametrization of Bosted and Christy \cite{Bosted:2007xd}
	in the resonance region, and a version of the ALLM fit published by the HERMES Collaboration \cite{Airapetian:2011nu} for the continuum
	region. It also uses information on the longitudinal structure function from SLAC \cite{Abe:1998ym}. As the fit is constructed closely following
	the LUXqed work Ref.\cite{Manohar:2017eqh}, we call this fit LUX-like.
	
	\item a Vector-Meson-Dominance model inspired fit of $F_2$ proposed in \cite{SU} at low $Q^2$, which is completed by the same NNLO MSTW structure function as above 
	at large $Q^2$. This fit is labelled SU for brevity.
\end{enumerate}

One can see from Table \ref{table:1} that the largest
inelastic-inelastic component is in all calculations systematically bigger 
than the elastic-elastic component, which gives the smallest contribution.
For the case of production of $e^+ e^-$ or $\mu^+ \mu^-$ via $\gamma \gamma$ fusion all components were
of the same size \cite{Luszczak:2015aoa}.

%------------------------------------------------------------------------------------
%\begin{table}
\begin{table}[tbp]
\centering
\begin{tabular}{|c|c|c|}
\hline
contribution               &  8 TeV  & 13 TeV  \\
\hline
        LUX-like             &      &          \\

$\gamma_{el} \gamma_{in}$  & 0.214 & 0.409   \\
$\gamma_{in} \gamma_{el}$  & 0.214 & 0.409   \\
$\gamma_{in} \gamma_{in}$  & 0.478 & 1.090   \\
\hline
      ALLM97 F2      &      &          \\

$\gamma_{el} \gamma_{in}$  & 0.197 & 0.318   \\
$\gamma_{in} \gamma_{el}$  & 0.197 & 0.318    \\
$\gamma_{in} \gamma_{in}$  & 0.289 & 0.701    \\
\hline
      SU F2 &      &          \\

$\gamma_{el} \gamma_{in}$  & 0.192 & 0.420    \\
$\gamma_{in} \gamma_{el}$  & 0.192 & 0.420    \\
$\gamma_{in} \gamma_{in}$  & 0.396 & 0.927    \\
\hline
    LUXqed collinear &      &          \\

%$\gamma_{el} \gamma_{in}$  & 0.141 & 0.288    \\
%$\gamma_{in} \gamma_{el}$  & 0.141 & 0.288   \\
$\gamma_{in+el}$ $ \gamma_{in+el}$  & 0.366 & 0.778    \\
\hline
    MRST04 QED collinear &      &          \\

$\gamma_{el} \gamma_{in}$  & 0.171 & 0.341    \\
$\gamma_{in} \gamma_{el}$  & 0.171 & 0.341    \\
$\gamma_{in} \gamma_{in}$  & 0.548 & 0.980    \\
\hline
    Elastic- Elastic&      &          \\
$\gamma_{el} \gamma_{el}$ (Budnev)  & 0.130 & 0.273   \\

$\gamma_{el} \gamma_{el}$  (DZ)  & 0.124 & 0.267  \\

\hline

\end{tabular}
\caption{Cross sections (in $p b$) for different contributions 
and different F2 structure functions: LUX, ALLM97 and SU, 
compared to the relevant collinear distributions with MRST04 QED and 
LUXqed distributions. The elastic-elastic contributions were obtained using fluxes from Refs. \cite{Budnev:1974de}, \cite{Drees:1988pp}.
}
\label{table:1}
\end{table}
%------------------------------------------------------------------------------------

We obtain cross sections of about 0.8--1 pb at $\sqrt{s}$ = 8 TeV 
and 1.5--1.8 pb at $\sqrt{s}$ = 13 TeV. 
This may be compared to 
41.1 $\pm$ 15.3 (stat) $\pm$ 5.8 (syst) $\pm$ 4.5 (lumi) pb (CMS \cite{CMS_inclusive})
and
54.4 $\pm$ 4.0 (stat) $\pm$ 3.9 (syst) $\pm$ 2.0 (lumi) pb (ATLAS \cite{ATLAS_inclusive})
measured (and extrapolated) at the LHC for $\sqrt{s}$ = 7 TeV.
This shows that the two-photon production constitutes about
2 \% of the total cross section. However, its relative contribution, as will 
be discussed below, increases with $M_{WW}$.

%--------------------------------------------------------
\subsection{One-dimensional distributions}
%--------------------------------------------------------

\begin{figure}[!htbp]
\begin{minipage}{0.47\textwidth}
 \centerline{\includegraphics[width=1.0\textwidth]{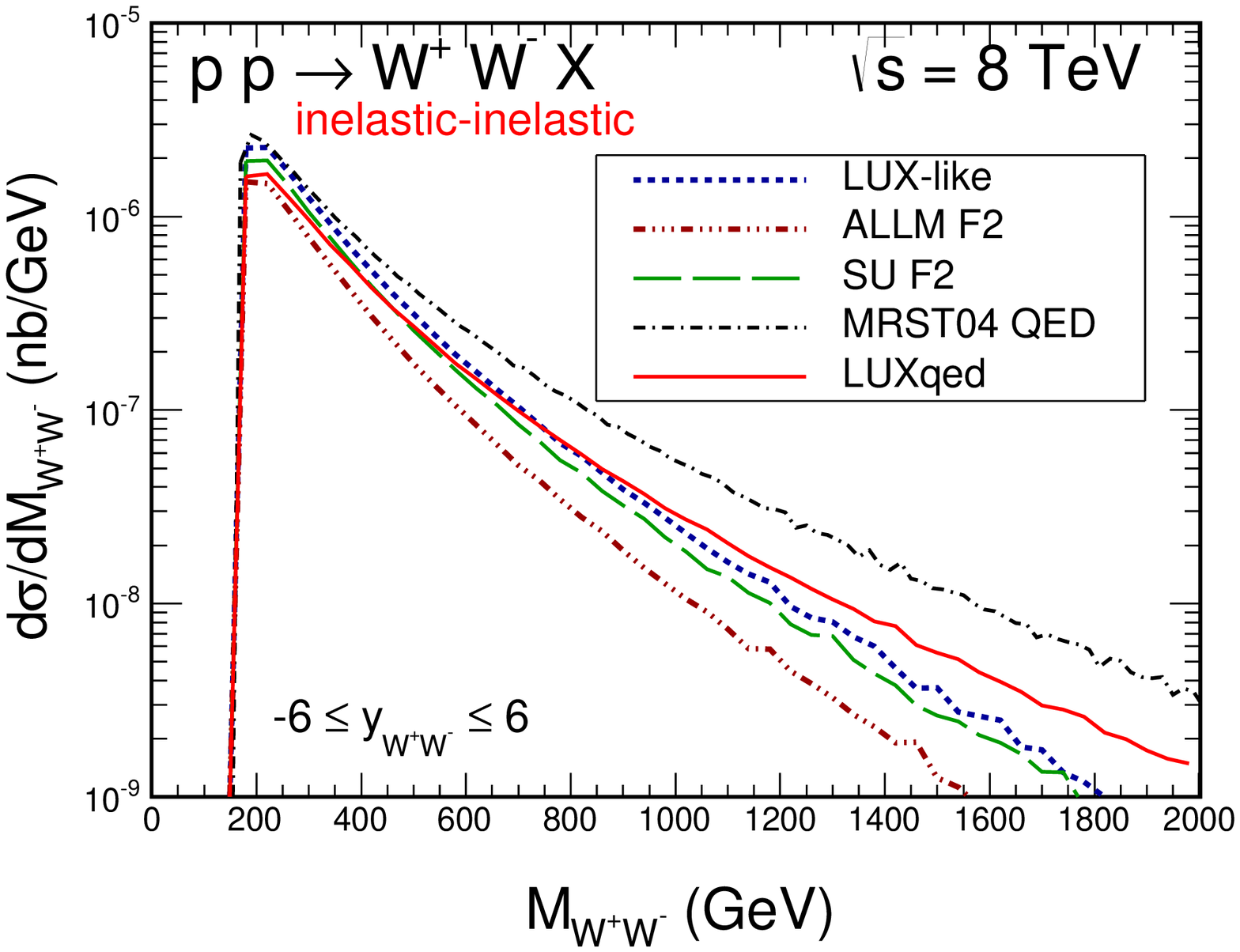}}
\end{minipage}
%\hspace{0.5cm}
\begin{minipage}{0.47\textwidth}
 \centerline{\includegraphics[width=1.0\textwidth]{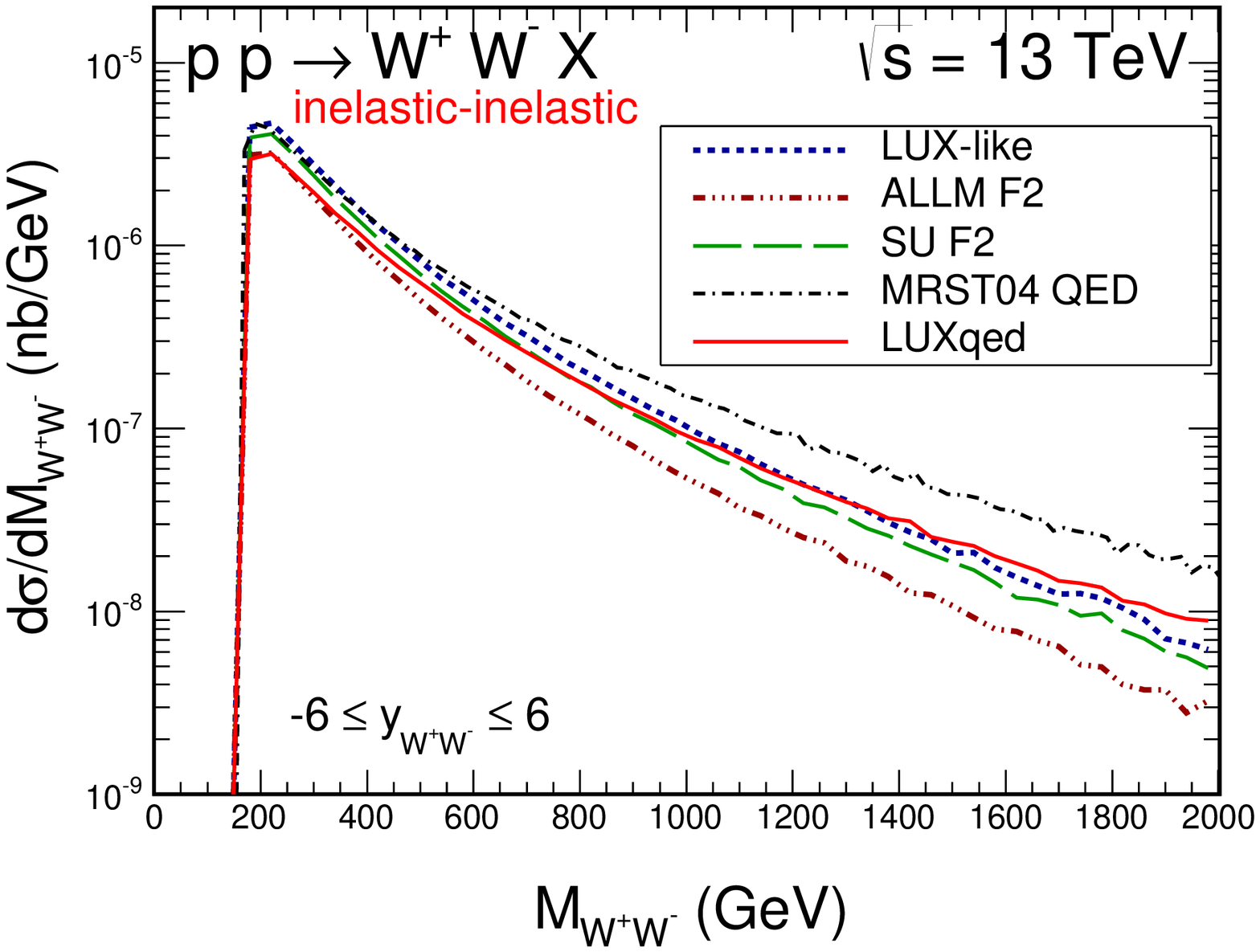}}
\end{minipage}
\caption{
\small
The inelastic-inelastic contribution to $W^+ W^-$ invariant mass 
distributions for different 
structure functions: LUX-like, ALLM97 , SU
compared to the relevant collinear distributions: MRST04 QED, LUXqed.
The left panel shows results for $\sqrt{s}$ = 8 TeV, while 
the right panel shows results for $\sqrt{s}$ = 13 TeV.
}
 \label{fig:dsig_dMWW_ineine}
\end{figure}

In Fig.\ref{fig:dsig_dMWW_ineine} we show invariant mass distributions
for $\sqrt{s}$ = 8 TeV (left panel) and $\sqrt{s}$ = 13 TeV (right
panel). The calculations have been performed with different
parametrizations of structure functions including the LUX-like one.
There are large uncertainties in the region of large invariant masses.
The uncertainties become smaller for larger $\sqrt{s}$.
We will compare to Ref.\cite{LSR2015}, i.e. to result of collinear calculations with
the rather old MRST04 QED set \cite{Martin:2004dh} (dash-dotted line). 
The new results should be regarded as an update of
the older results in \cite{LSR2015}. 

%-----------------------------------------------------------------------------
%------------------------------------------------------------------------------

\begin{figure}[!htbp]
\begin{minipage}{0.47\textwidth}
 \centerline{\includegraphics[width=1.0\textwidth]{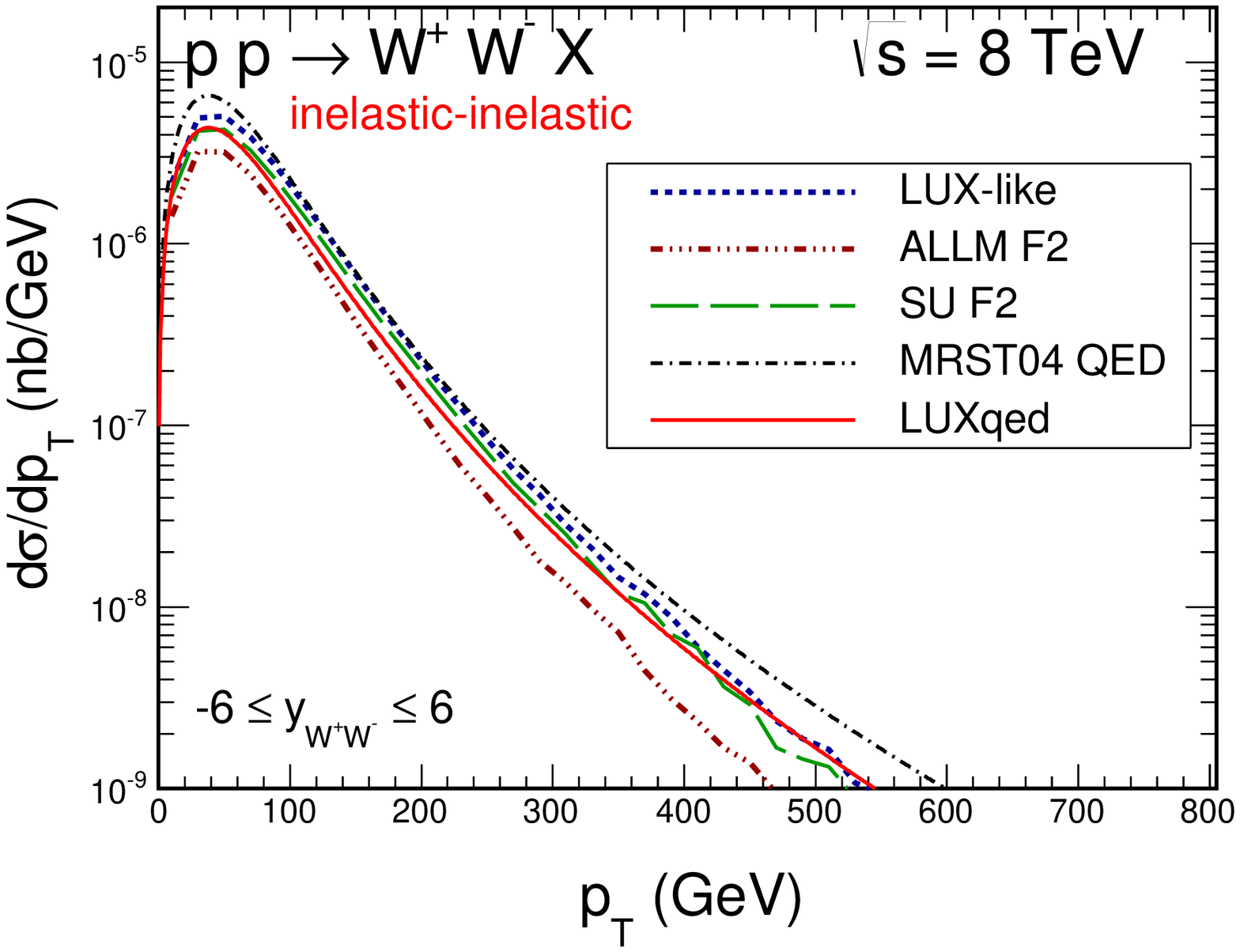}}
\end{minipage}
%\hspace{0.5cm}
\begin{minipage}{0.47\textwidth}
 \centerline{\includegraphics[width=1.0\textwidth]{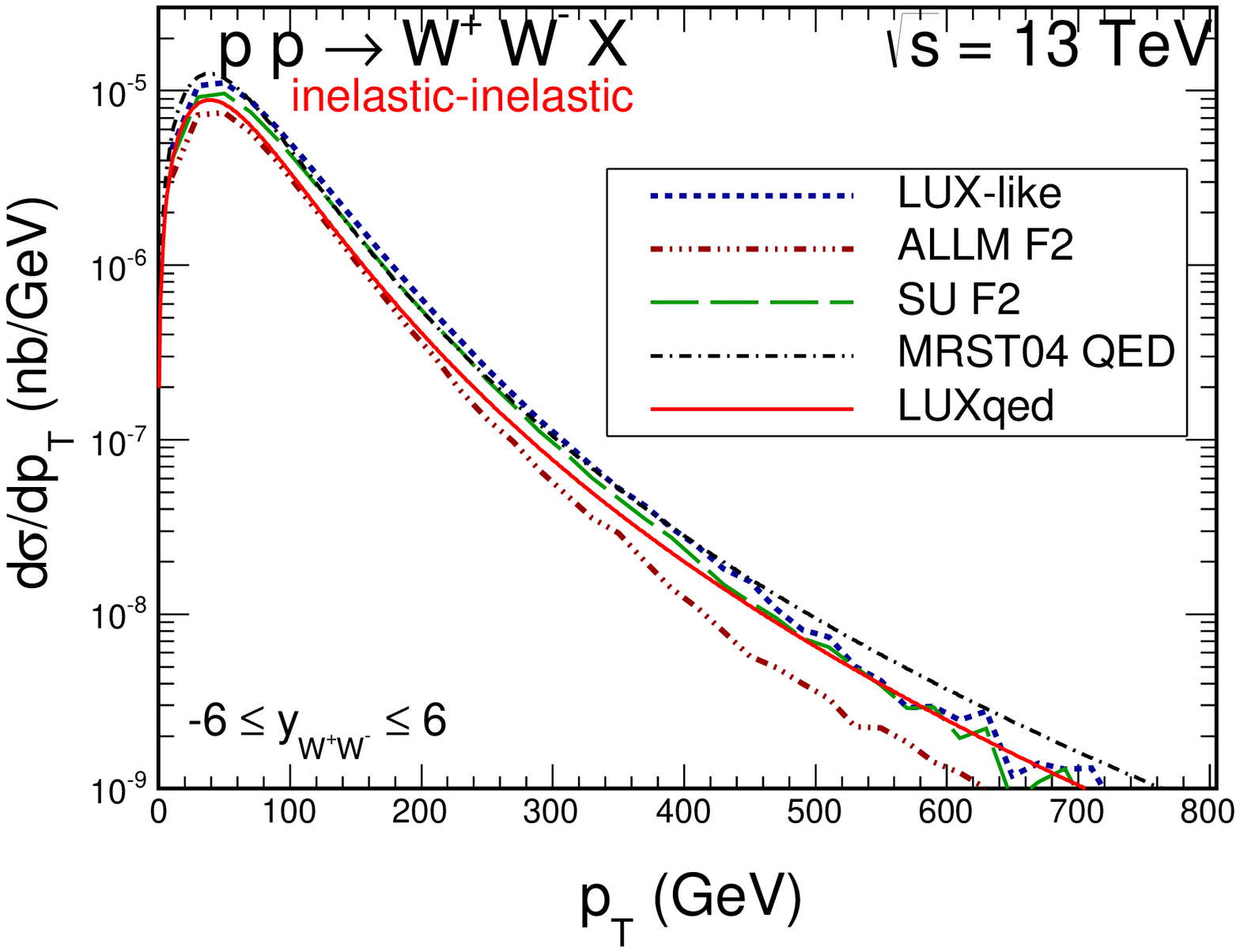}}
\end{minipage}
\caption{
\small
Transverse momentum distribution of $W^+$ or $W^-$ bosons 
for different structure functions: LUX-like, ALLM97, SU 
compared to the relevant collinear distributions: MRST04 QED, LUXqed.
The left panel shows results for $\sqrt{s}$ = 8 TeV, while 
the right panel shows results for $\sqrt{s}$ = 13 TeV.
}
 \label{fig:dsig_dpt}
\end{figure}

The distribution in transverse momentum of a $W$ boson is shown in 
Fig.\ref{fig:dsig_dpt}. At low transverse momenta there is a relatively 
small theoretical uncertainty. The result obtained with our LUX-like
structure function should be considered as our best estimate.

%-----------------------------------------------------------------------------
%------------------------------------------------------------------------------

\begin{figure}[!htbp]
\begin{minipage}{0.47\textwidth}
 \centerline{\includegraphics[width=1.0\textwidth]{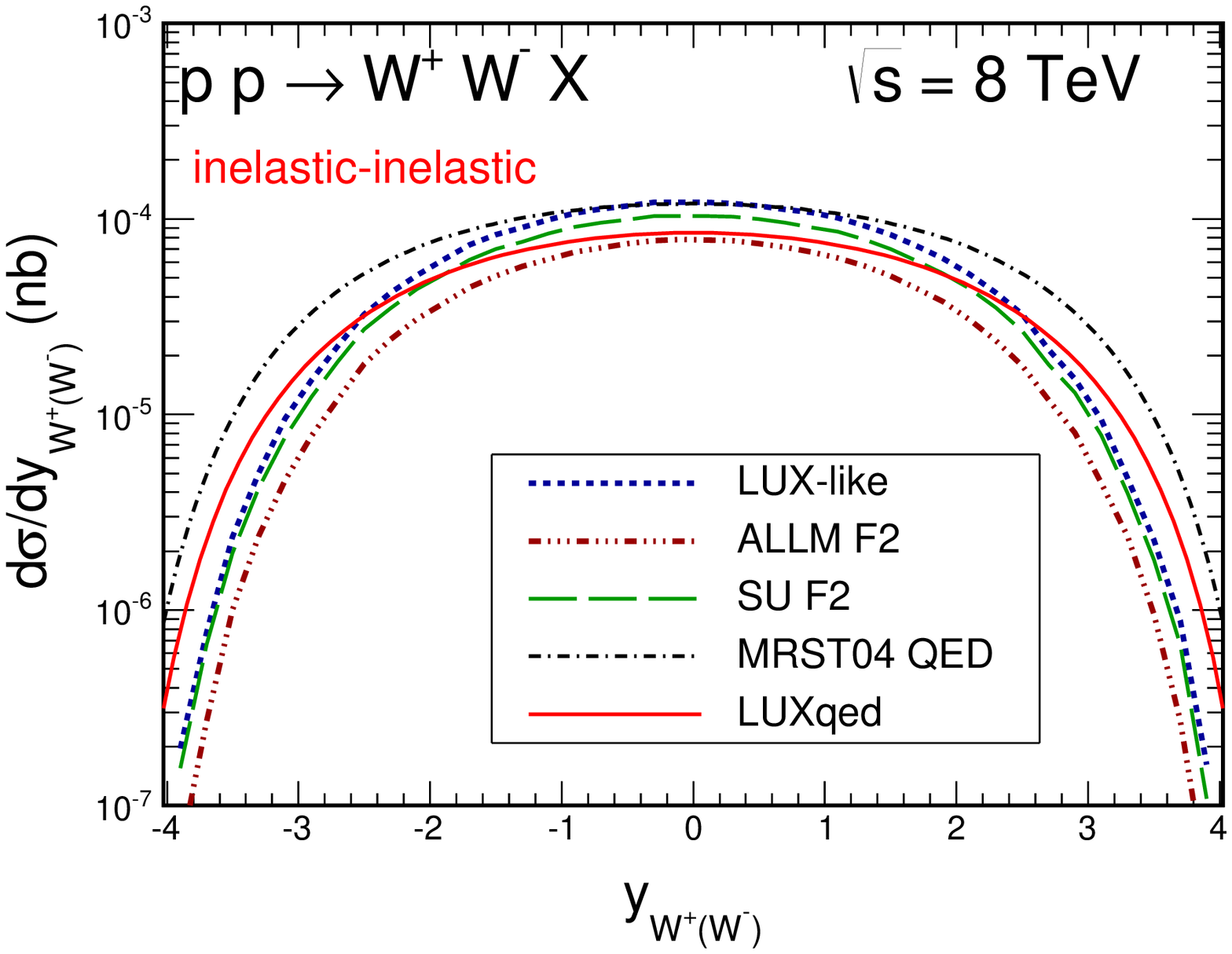}}
\end{minipage}
%\hspace{0.5cm}
\begin{minipage}{0.47\textwidth}
 \centerline{\includegraphics[width=1.0\textwidth]{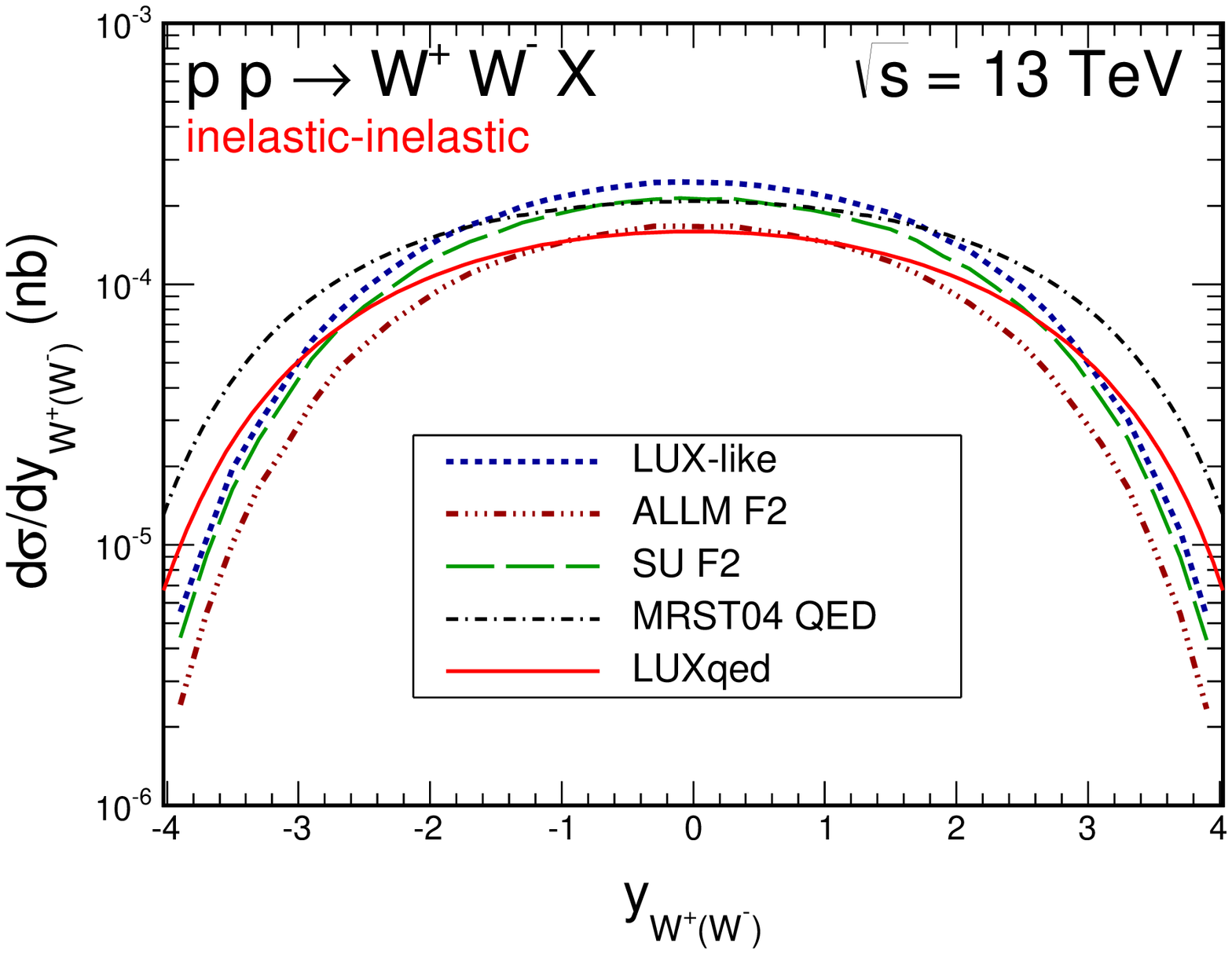}}
\end{minipage}
\caption{
\small
Rapidity distribution of $W^+$ or $W^-$ bosons
for different structure functions: LUX-like, ALLM97, SU.
The left panel shows results for the 
$\sqrt{s}$ = 8 TeV, while the right panel shows results for 
$\sqrt{s}$ = 13 TeV.
%---------------------------------------------------------------------
}
\label{fig:dsig_dy}
\end{figure}

For completeness we show also rapidity distributions of $W^+ / W^-$
bosons in Fig.\ref{fig:dsig_dy}.
The distribution in collinear approach extends to much larger
rapidities, especially for $\sqrt{s}$ = 13 TeV.

%-----------------------------------------------------------------------------
%---------------------------------------------------------------

\begin{figure}[!htbp]
\begin{minipage}{0.47\textwidth}
 \centerline{\includegraphics[width=1.0\textwidth]{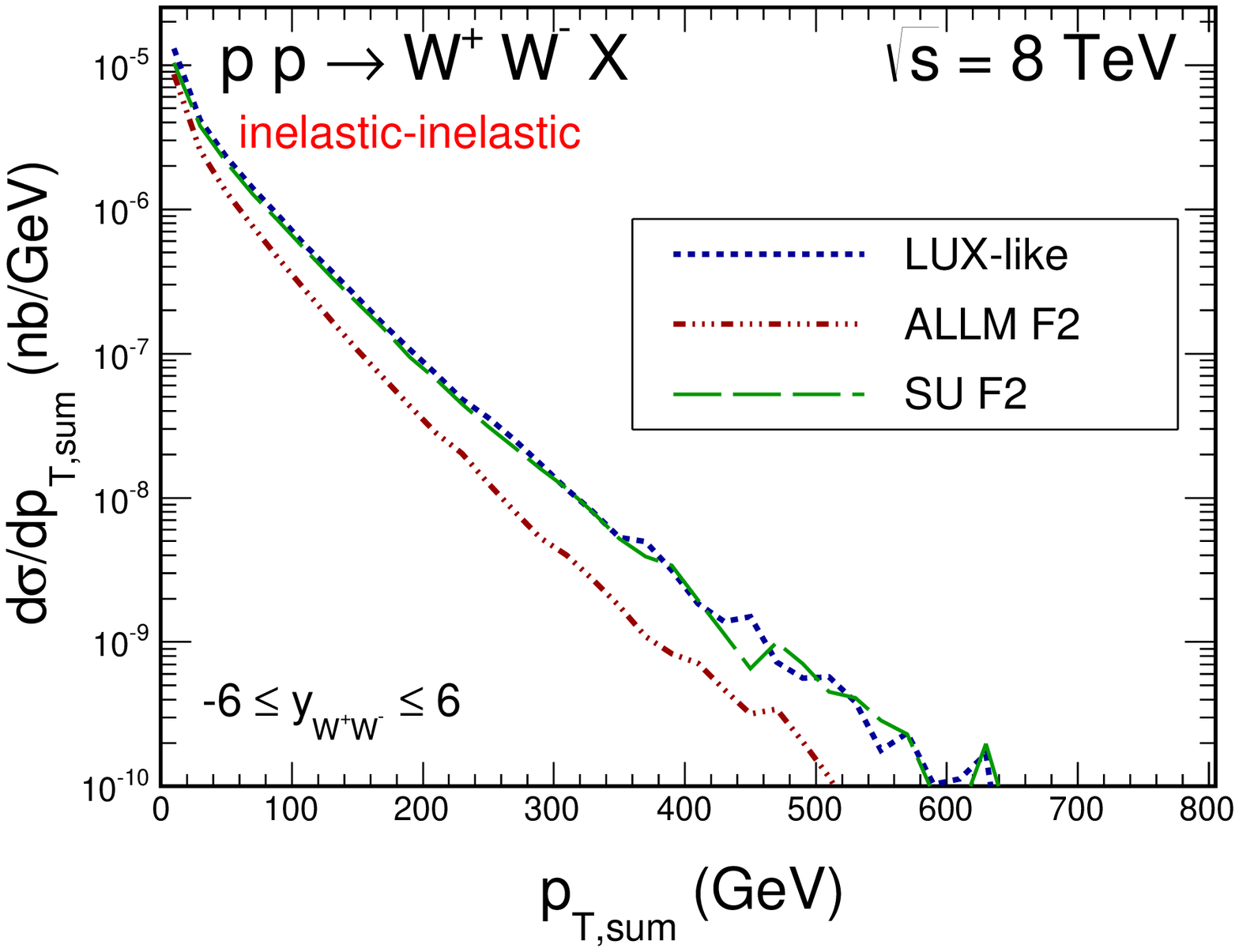}}
\end{minipage}
%\hspace{0.5cm}
\begin{minipage}{0.47\textwidth}
 \centerline{\includegraphics[width=1.0\textwidth]{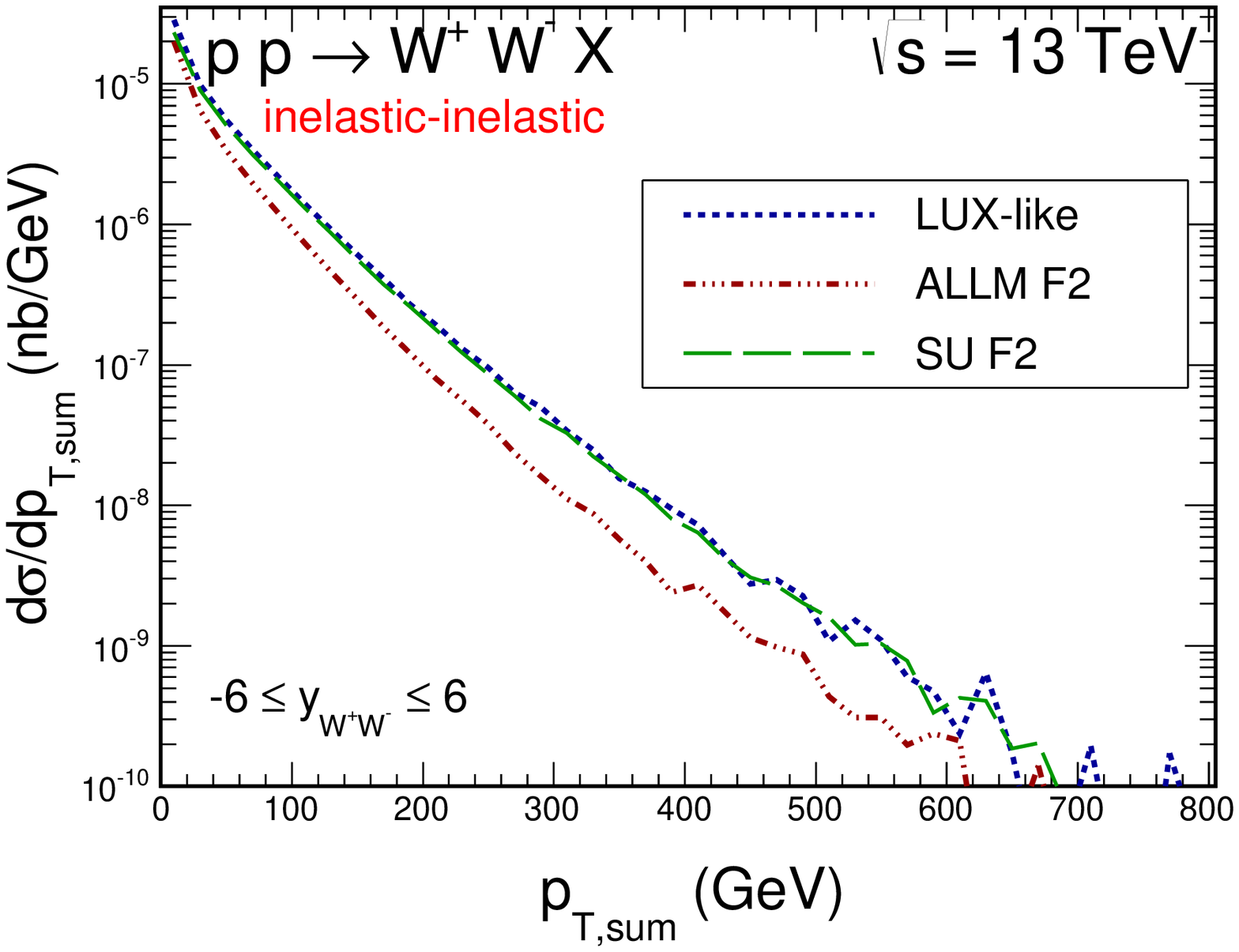}}
\end{minipage}
\caption{
\small
Transverse momentum distribution of $W^+W^-$  bosons 
for different structure functions: LUX-like, ALLM97, SU.
The left panel shows results for
$\sqrt{s}$ = 8 TeV, while the right panel shows results for 
$\sqrt{s}$ = 13 TeV.
}
 \label{fig:dsig_dptsum}
\end{figure}

In Fig.\ref{fig:dsig_dptsum} we show distribution in transverse momentum
of the $W^+ W^-$ pair, $p_{T,sum}$. Quite large pair transverse momenta are possible.
In contrast in leading-order using collinear partons, the corresponding
distribution is just a Dirac delta function at $p_{T,sum}$ = 0. 
The $k_T$-factorization approach should be therefore here a much better 
approach.
This distribution is, however, a bit academic as in practice one
measures only charged leptons and the neutrinos escape experimental observation, 
but the figure demonstrates theoretical preference of the $k_T$-factorization
approach over the collinear approach. The nonvanishing pair transverse momentum can 
influence the transverse momentum distributions of associated
leptons (usually $\mu^+ e^-$ or $\mu^- e^+$) when it is large.
This effect will be discussed elsewhere.

%-----------------------------------------------------------------------------
%------------------------------------------------------------------------------

\begin{figure}[!htbp]
\begin{minipage}{0.47\textwidth}
 \centerline{\includegraphics[width=1.0\textwidth]{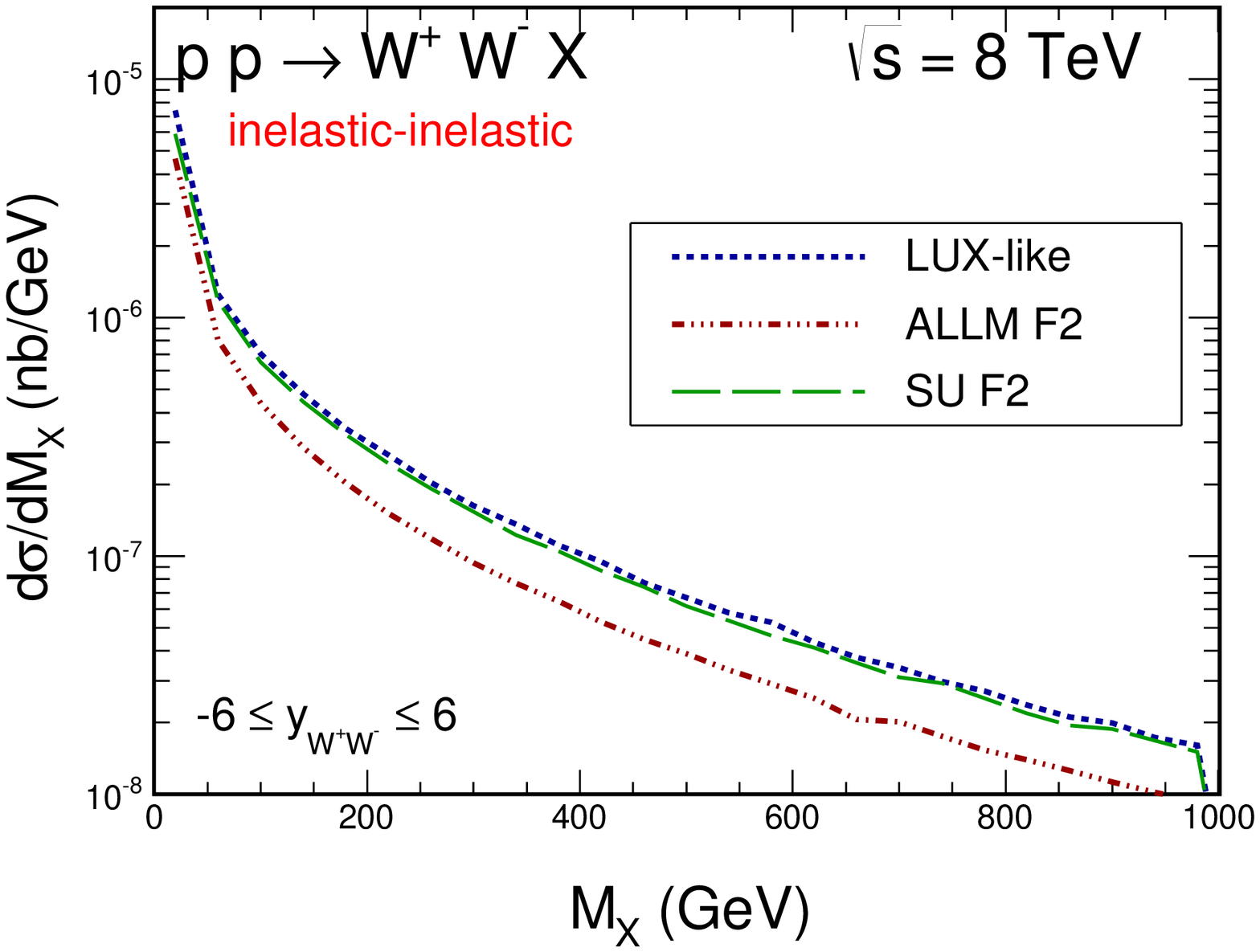}}
\end{minipage}
%\hspace{0.5cm}
\begin{minipage}{0.47\textwidth}
 \centerline{\includegraphics[width=1.0\textwidth]{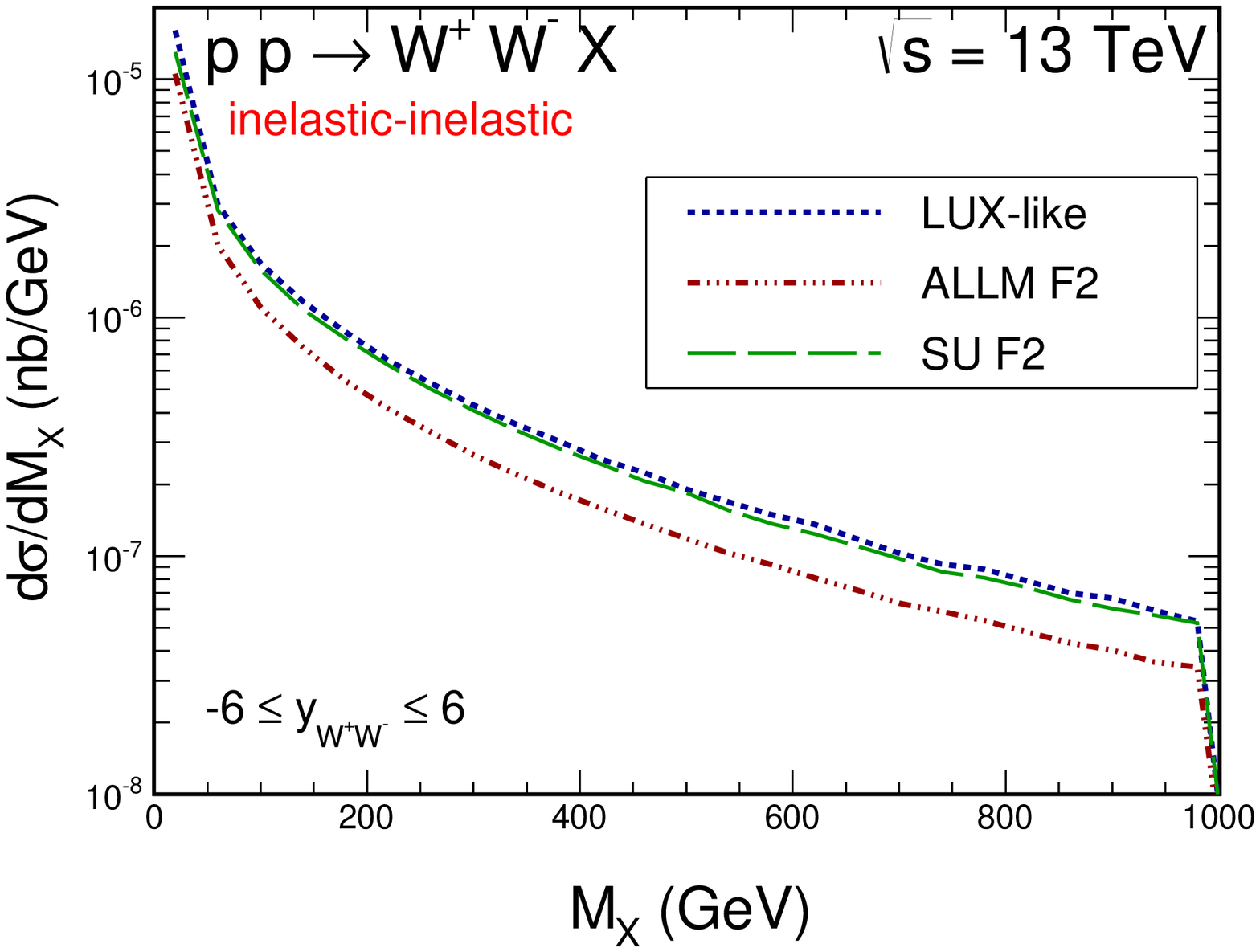}}
\end{minipage}
\caption{
\small
Missing mass distributions for inelastic-inelastic 
photon-photon contributions for different parametrizations of the
structure functions as explained inside the figures for two energies: 
$\sqrt{s}$ = 8 TeV (left panel) and $\sqrt{s}$ = 13 TeV (right panel).
}
 \label{fig:dsig_dMrem}
\end{figure}

Our approach also goes beyond \cite{Manohar:2016nzj,Manohar:2017eqh} in that it 
allows us to obtain the distribution of the mass of the proton remnant(s). 
These distributions are shown in Fig.\ref{fig:dsig_dMrem}. 
Quite large masses of the remnant system are generated. 
Notice, that the larger is the invariant mass, the smaller is the rapidity gap
from the proton remnant to the $WW$ system. 
Detailed studies of this effect require a hadronisation of the remnant 
system, which goes beyond the scope of the present paper.

%-----------------------------------------------------------------------------
%------------------------------------------------------------------------------

\begin{figure}[!htbp]
\begin{minipage}{0.47\textwidth}
 \centerline{\includegraphics[width=1.0\textwidth]{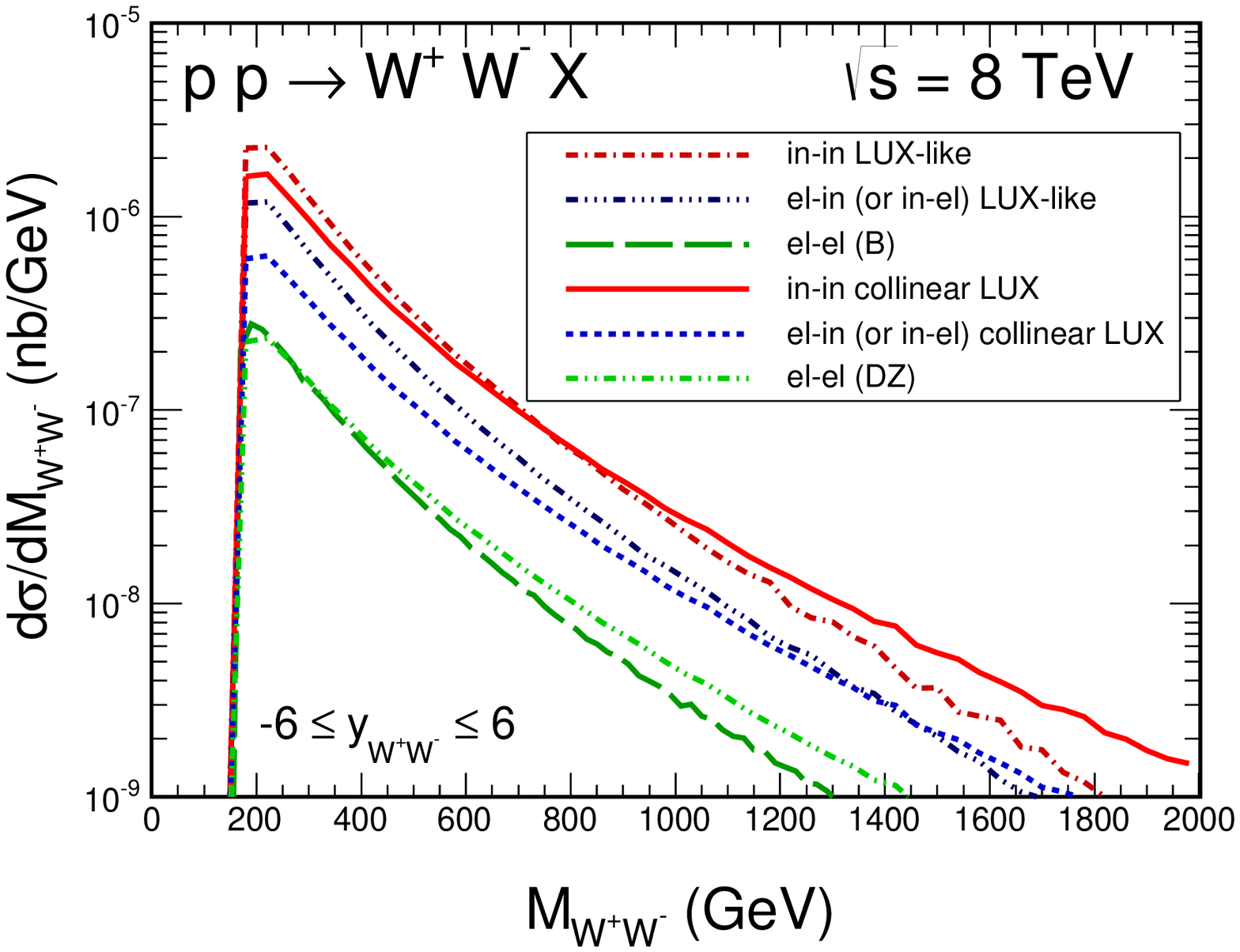}}
\end{minipage}
%\hspace{0.5cm}
\begin{minipage}{0.47\textwidth}
 \centerline{\includegraphics[width=1.0\textwidth]{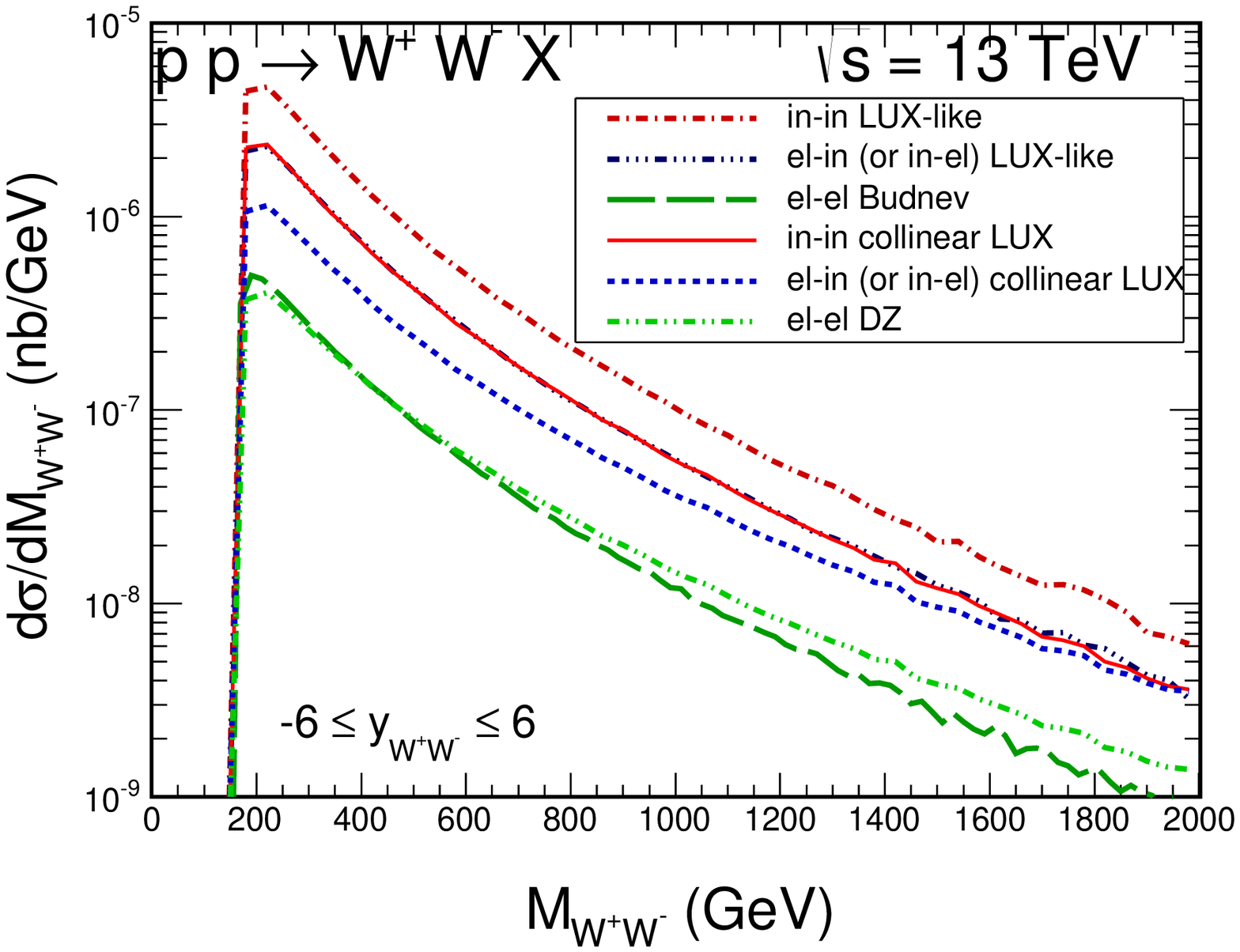}}
\end{minipage}
\caption{
\small
The inelastic-inelastic, elastic-inelastic, inelastic-elastic and 
elastic-elastic  contributions to $W^+ W^-$ invariant mass 
distributions for the $k_T$-factorization approach with 
the LUX-like structure function compared to the relevant 
distribution for collinear approach with similar structure
function LUXqed.
The left panel shows results for $\sqrt{s}$ = 8 TeV, while 
the right panel shows results for $\sqrt{s}$ = 13 TeV.
}
 \label{fig:dsig_dM_all}
\end{figure}

Now we shall compare results corresponding to different diagrams 
shown in Fig.\ref{fig:new_diagrams}. 
We start by showing distributions in invariant mass 
(see Fig.\ref{fig:dsig_dM_all}).
The inelastic contributions (inelastic-inelastic, inelastic-elastic
or elastic-inelastic) are larger than the purely elastic
(elastic-elastic) contribution. For reference we show distributions
in the collinear approach with the LUXqed structure function parametrization.

%-----------------------------------------------------------------------------
%------------------------------------------------------------------------------

\begin{figure}[!htbp]
\begin{minipage}{0.47\textwidth}
 \centerline{\includegraphics[width=1.0\textwidth]{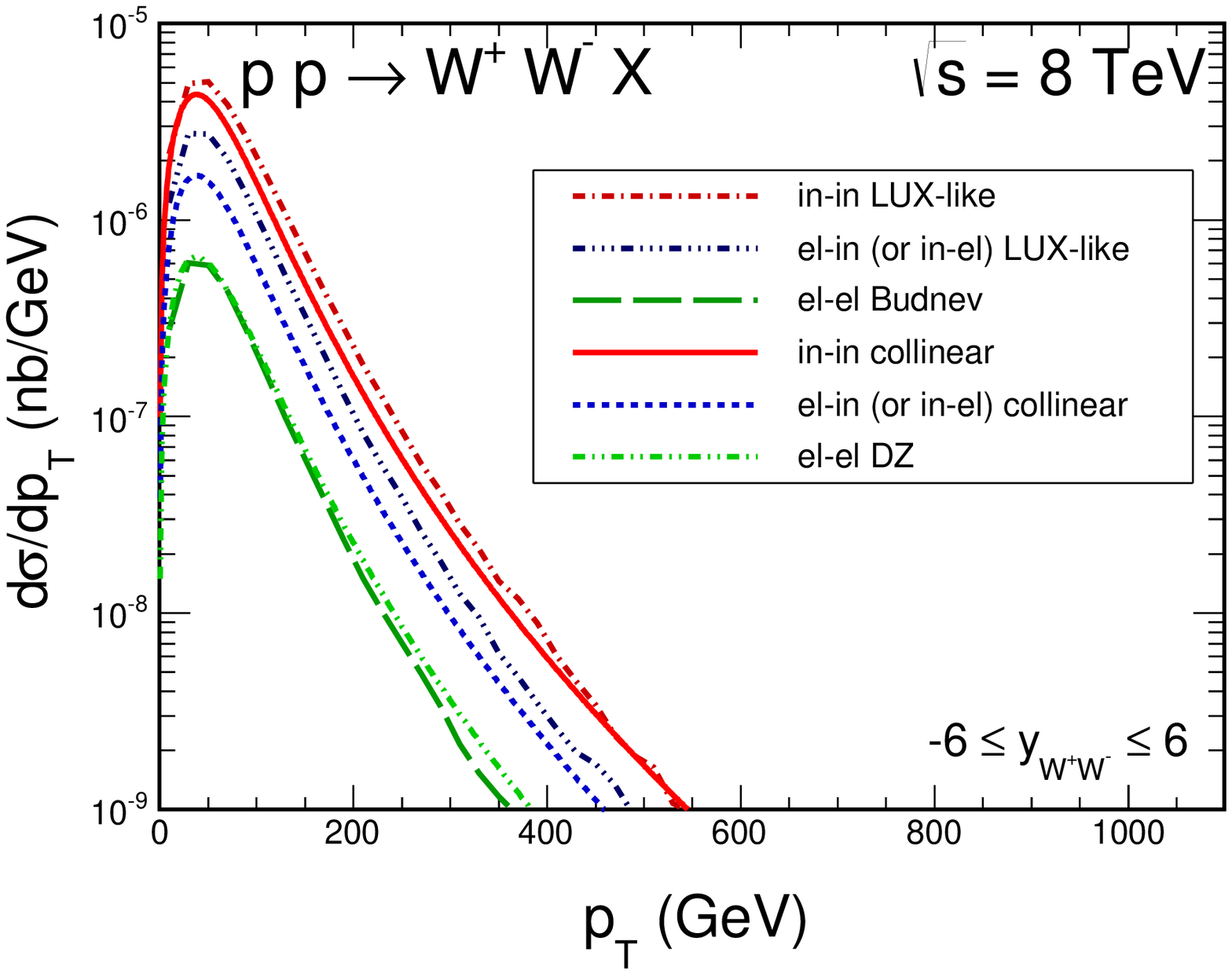}}
\end{minipage}
%\hspace{0.5cm}
\begin{minipage}{0.47\textwidth}
 \centerline{\includegraphics[width=1.0\textwidth]{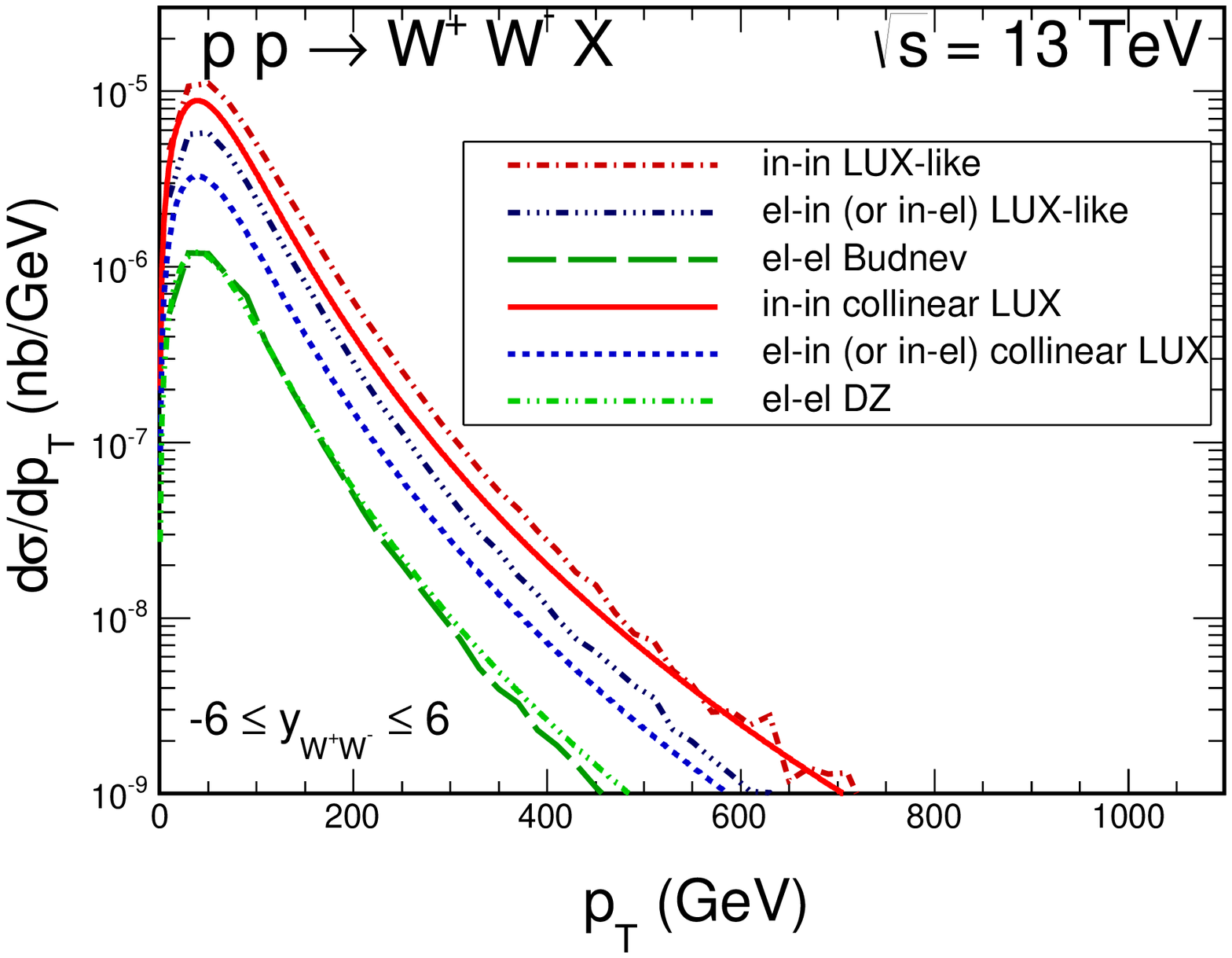}}
\end{minipage}
\caption{
\small
Transverse momentum distribution of $W^+$ or $W^-$ bosons for 
the inelastic-inelastic, elastic-inelastic, inelastic-elastic and 
elastic-elastic contributions for LUX-like structure function
compared to the relevant distribution for collinear approach with the 
LUXqed.
The left panel shows results for $\sqrt{s}$ = 8 TeV, while 
the right panel shows results for $\sqrt{s}$ = 13 TeV.
}
 \label{fig:dsig_dpt_all}
\end{figure}

In Fig.\ref{fig:dsig_dpt_all} we compare transverse momentum
distributions for all components of Fig.\ref{fig:new_diagrams}.
Similar slopes are obtained for different components, while the
corresponding cross sections are different.

%-----------------------------------------------------------------------------
%------------------------------------------------------------------------------

A similar result for the pair transverse momentum distribution is shown in 
Fig.\ref{fig:dsig_dpt_all}. The distribution for 
the inelastic-inelastic contribution is broader than that for
elastic-inelastic or inelastic-elastic component. The elastic-elastic
contribution gives very narrow distribution compared to the two other
components.

%-----------------------------------------------------------------------------
\begin{figure}[!htbp]
\begin{minipage}{0.47\textwidth}
 \centerline{\includegraphics[width=1.0\textwidth]{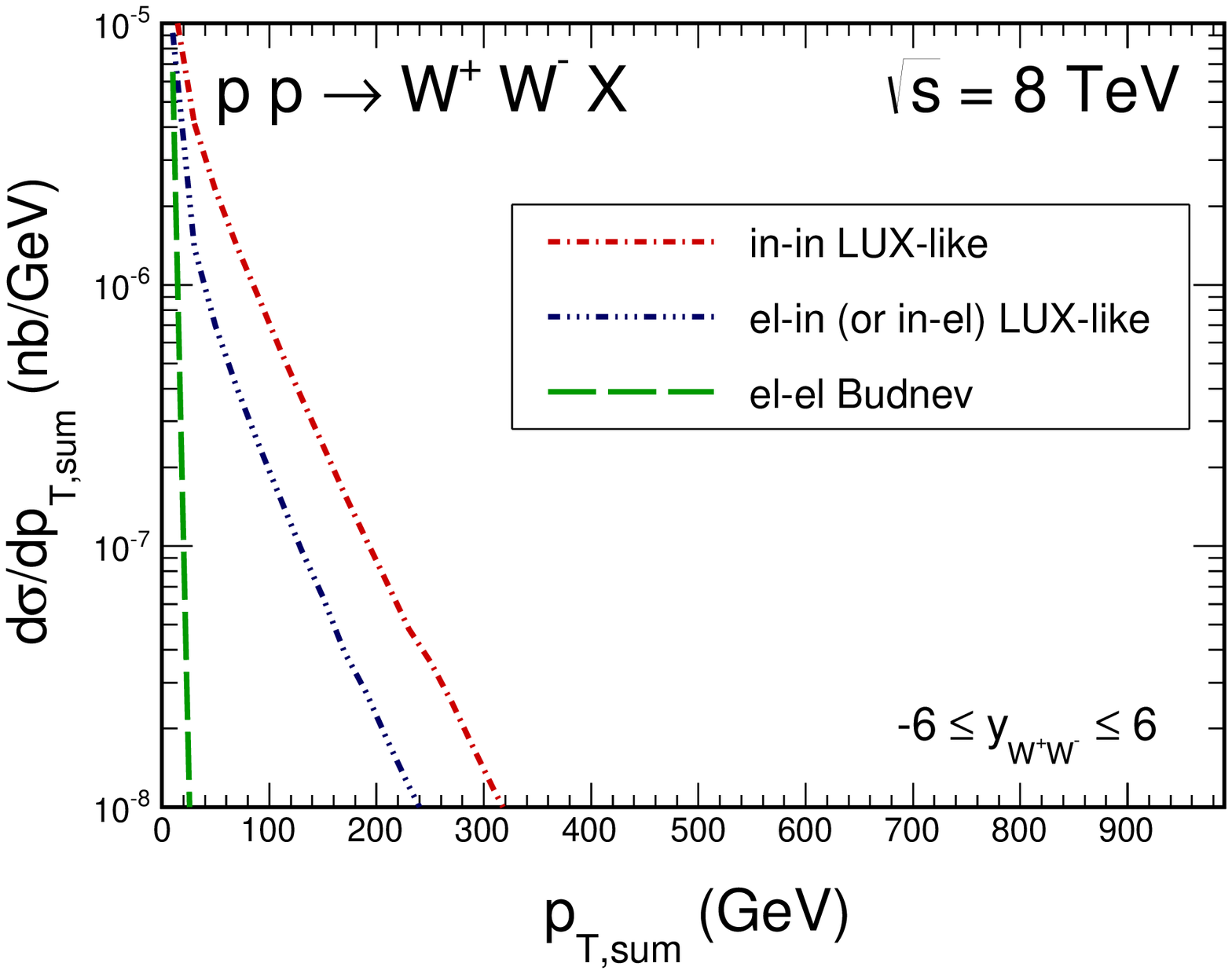}}
\end{minipage}
%\hspace{0.5cm}
\begin{minipage}{0.47\textwidth}
 \centerline{\includegraphics[width=1.0\textwidth]{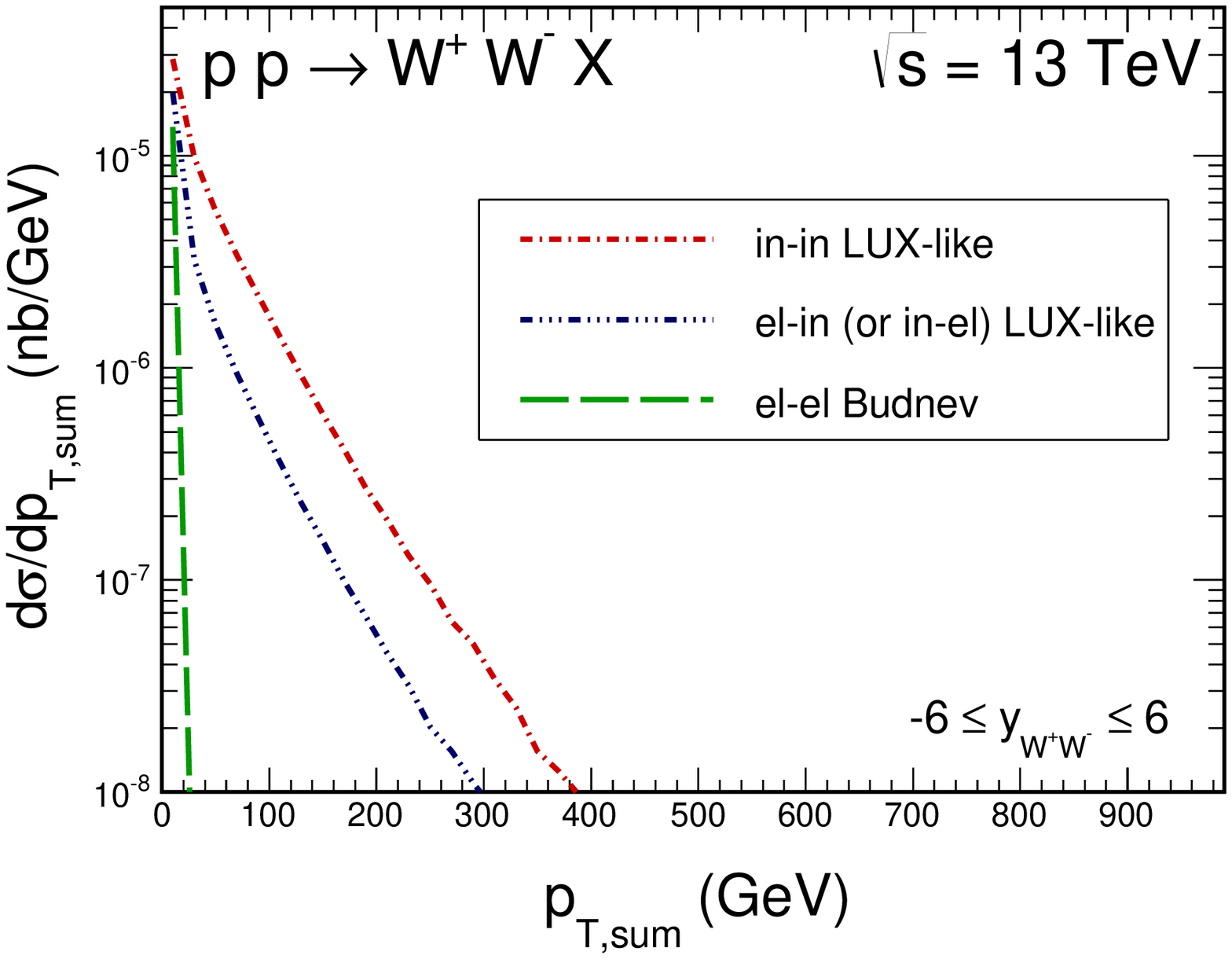}}
\end{minipage}
\caption{
\small
Distribution in transverse momentum of the $W^+ W^-$ pairs for 
the inelastic-inelastic, elastic-inelastic, inelastic-elastic
and elastic-elastic  contributions for the LUX-like structure function.
The left panel shows results for
$\sqrt{s}$ = 8 TeV, while the right panel shows results for 
$\sqrt{s}$ = 13 TeV.
}
 \label{fig:dsig_dy_all}
\end{figure}
%------------------------------------------------------------------------------

\begin{figure}[!htbp]
\begin{minipage}{0.47\textwidth}
 \centerline{\includegraphics[width=1.0\textwidth]{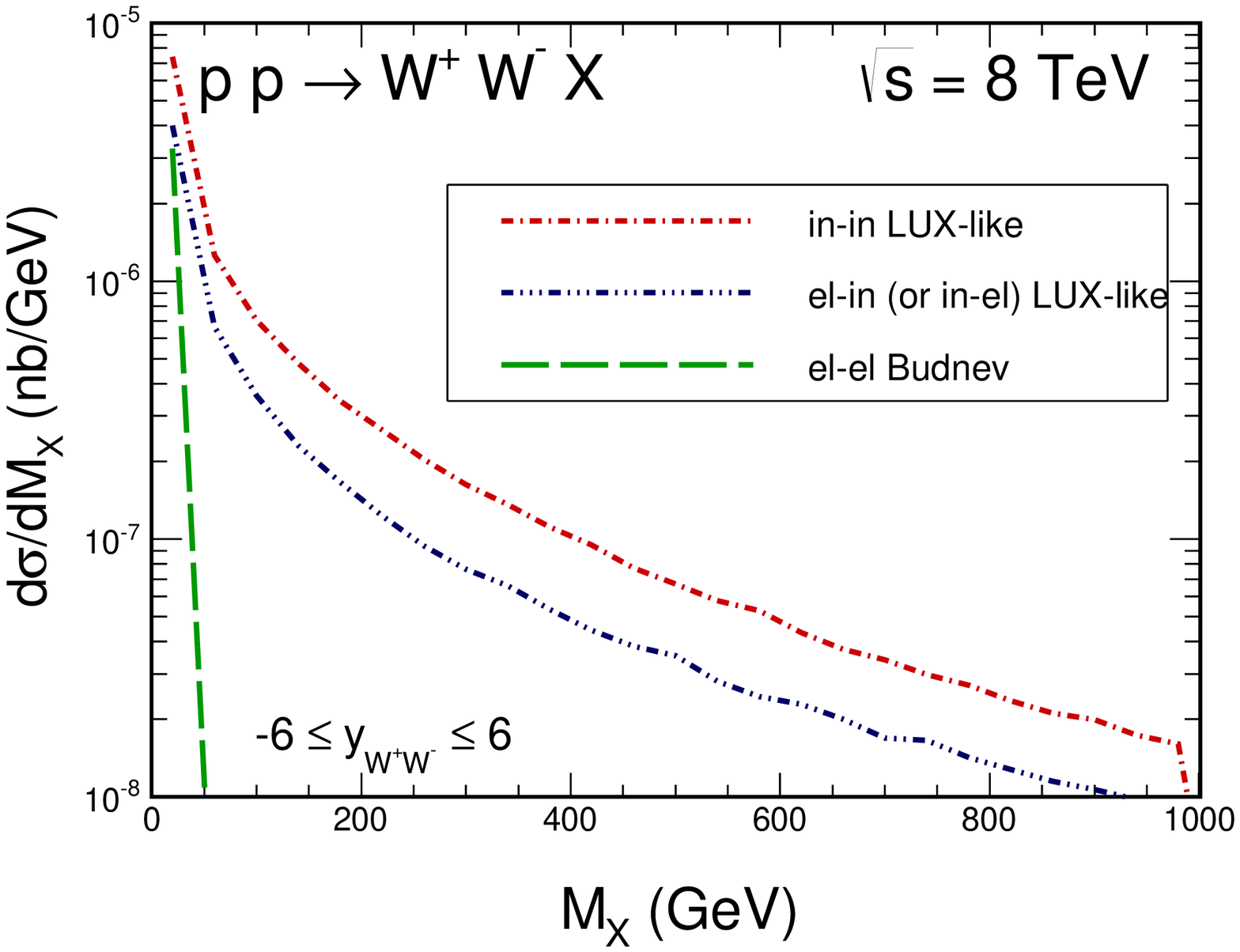}}
\end{minipage}
%\hspace{0.5cm}
\begin{minipage}{0.47\textwidth}
 \centerline{\includegraphics[width=1.0\textwidth]{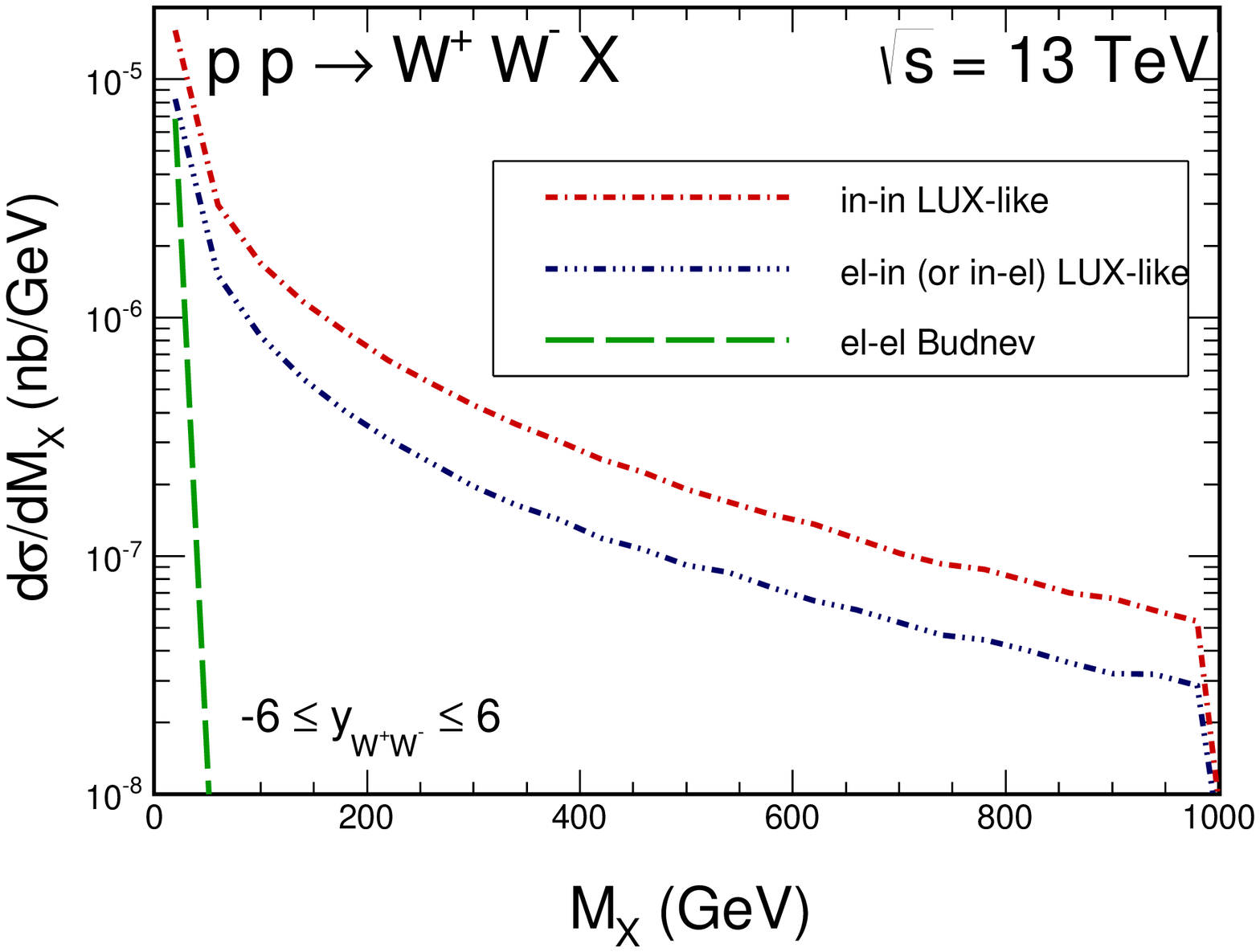}}
\end{minipage}
\caption{
\small
Missing mass distributions for the inelastic-inelastic,
elastic-inelastic, inelastic-elastic and elastic-elastic 
contributions for the LUX-like structure function.
The left panel shows results for W = 8 TeV, while the right panel 
shows results for W = 13 TeV.
}
 \label{fig:dsig_dMX}
\end{figure}

The missing mass distributions for different components are shown
in Fig.\ref{fig:dsig_dMX}. The shape for the elastic-inelastic and
inelastic-elastic is the same as that for inelastic-inelastic component.
The one for the elastic-elastic contribution is just the Dirac delta
function at $M_X = M_Y = m_p$. We shall return to the issue whether
the distributions in $M_X$ and $M_Y$ for the inelastic-inelastic
component are correlated when discussing two-dimensional distributions
of correlation character.

%-----------------------------------------------------------------------------
%------------------------------------------------------------------------------

%-----------------------------------------------
\subsection{Correlation observables}
%-----------------------------------------------

Now we shall proceed to two-dimensional distributions
of correlation character.

\begin{figure}[!h]
\begin{minipage}{0.3\textwidth}
 \centerline{\includegraphics[width=1.0\textwidth]{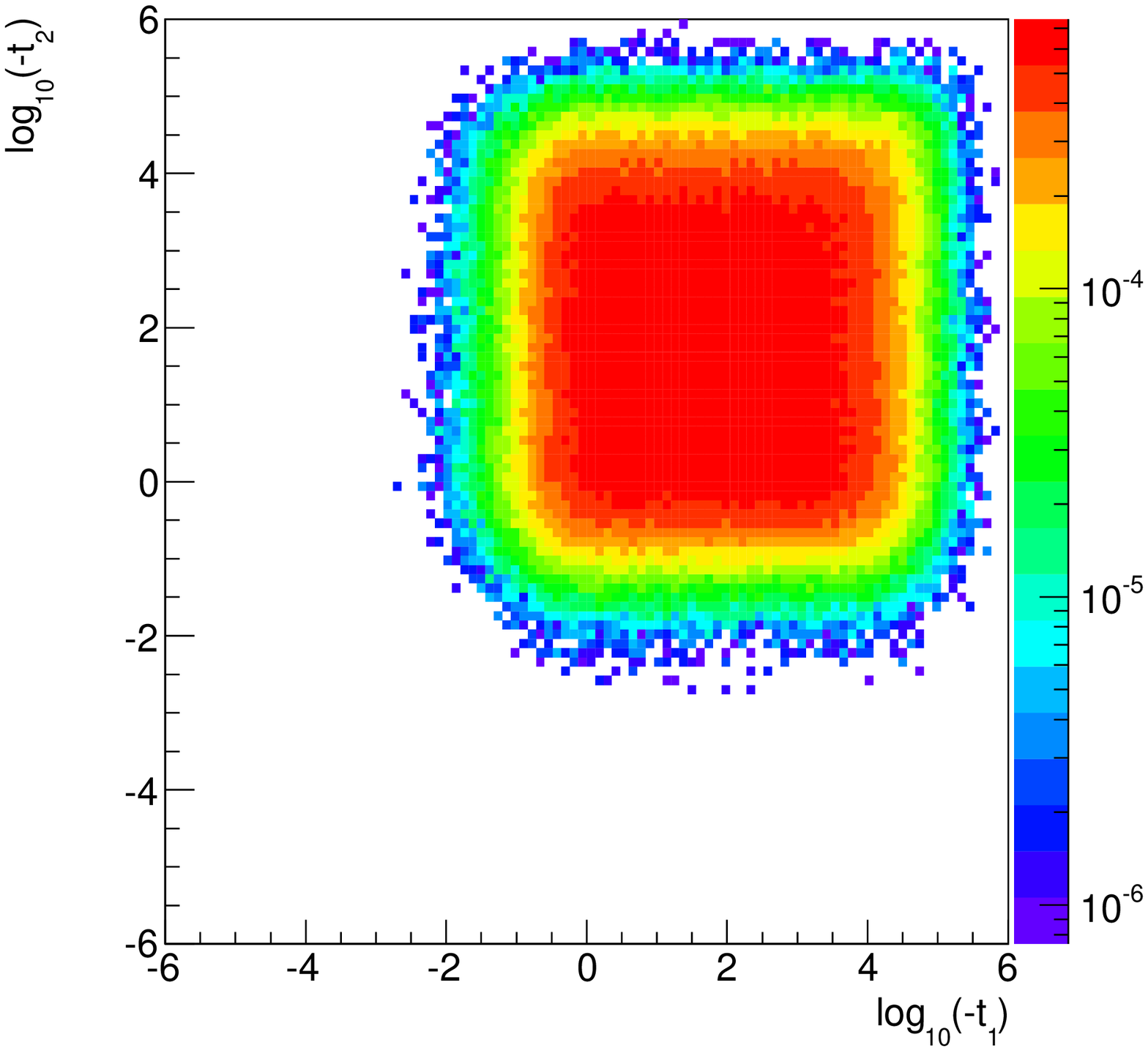}}
\end{minipage}
\hspace{0.2cm}
\begin{minipage}{0.3\textwidth}
 \centerline{\includegraphics[width=1.0\textwidth]{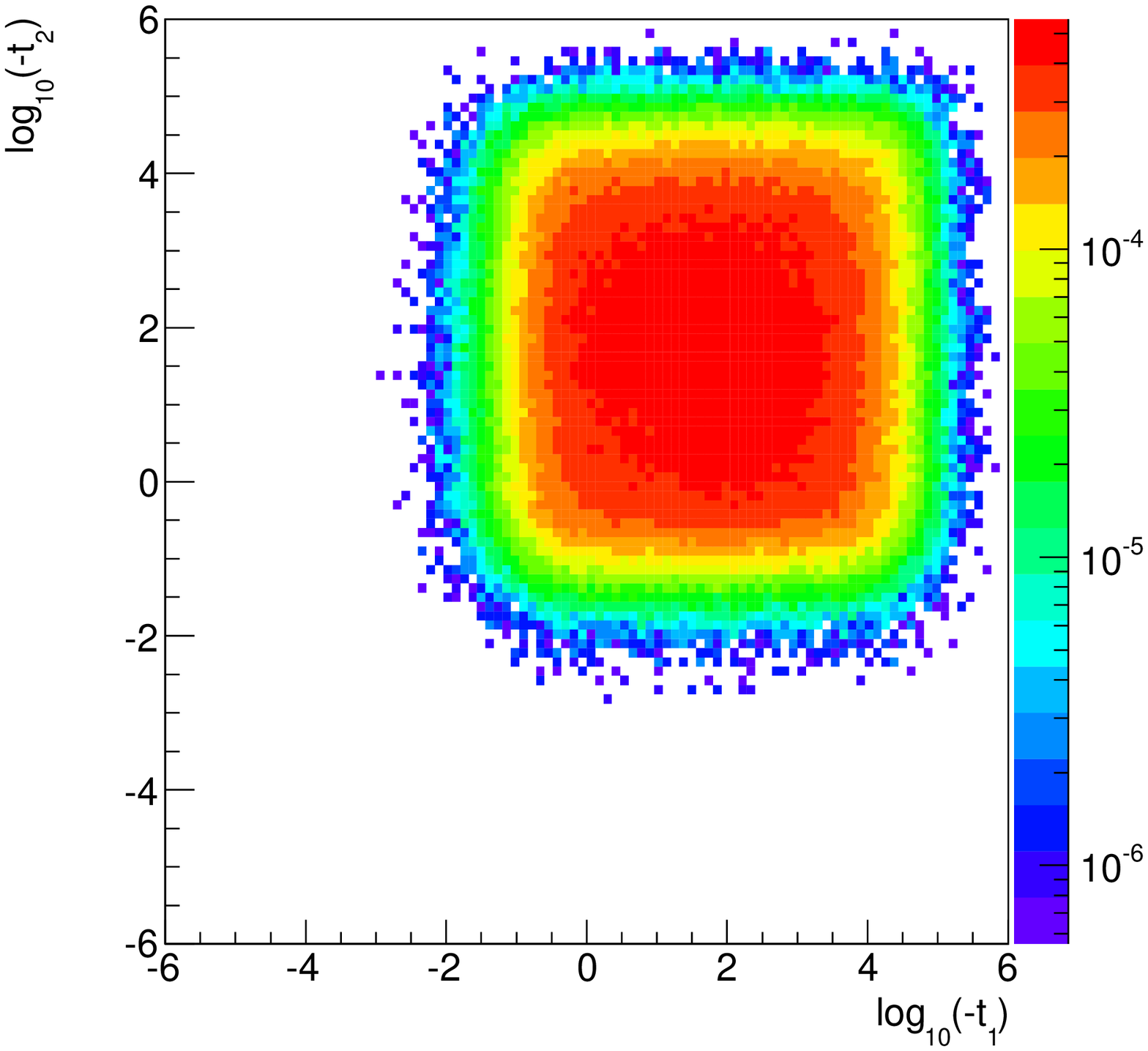}}
\end{minipage}
\hspace{0.2cm}
\begin{minipage}{0.3\textwidth}
 \centerline{\includegraphics[width=1.0\textwidth]{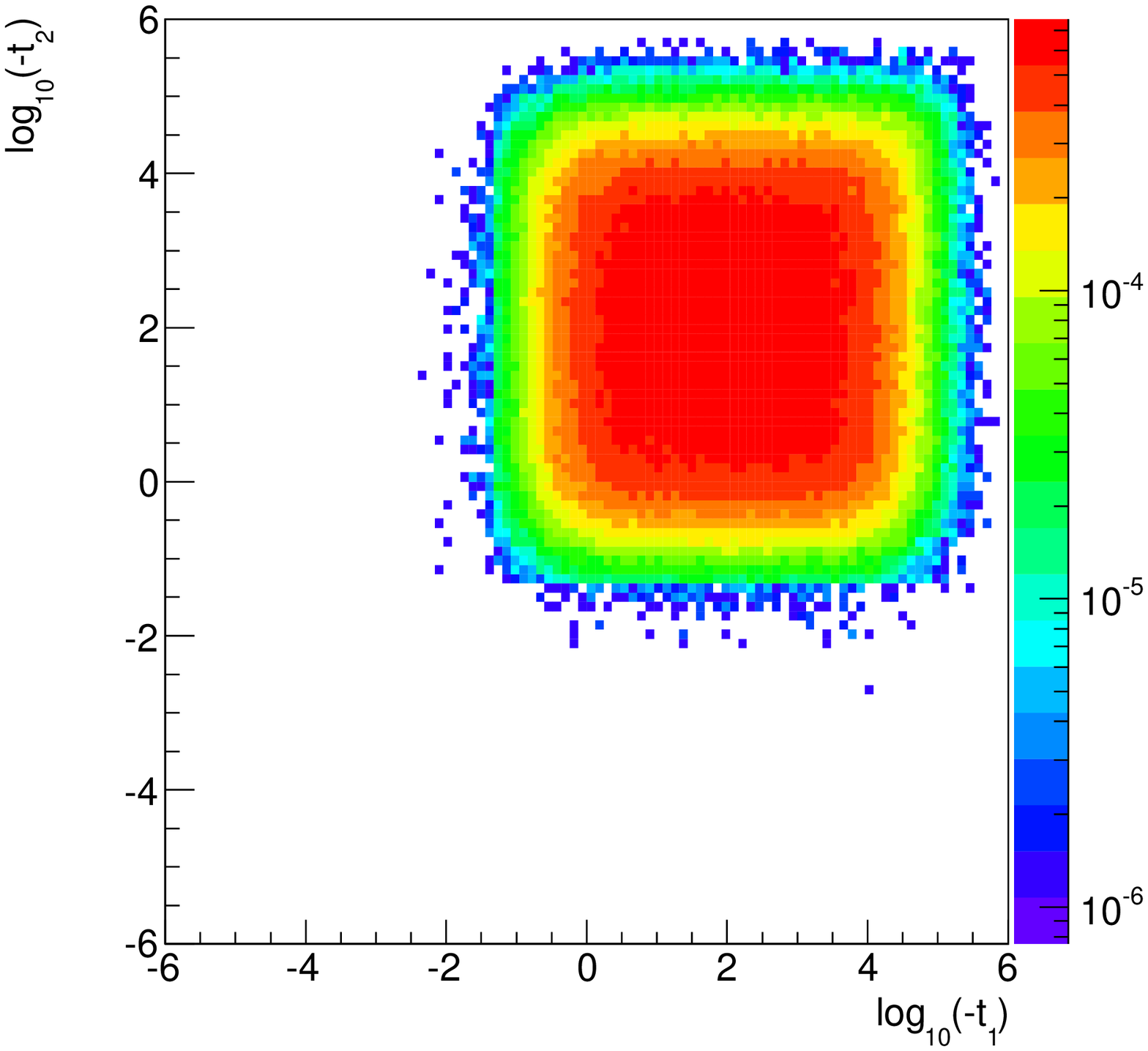}}
\end{minipage}

   \caption{Distributions for $Q_1^2 \times Q_2^2$ for different 
structure functions: LUX-like, ALLM97, SU
for $\sqrt{s}$ = 13 TeV.
}
\label{fig:dsig_dQ12dQ22}
\end{figure}

In the collinear approximation, the incoming photons are taken to be on-mass
shell, i.e. massless. How the situation changes in our approach
will be discussed in the following.
In Fig.\ref{fig:dsig_dQ12dQ22} we show distribution in
$Q_1^2 \times Q_2^2$ (please note logarithmic scales on both axes).
A plateau extending to $Q_1^2, Q_2^2 \sim$ 10$^4$ GeV 
can be seen. The result shows that collinear-factorization approach
could be far from being realistic for the $W^+ W^-$ production,
at least in some parts of the phase space.

%-----------------------------------------------------------------------%
%----------------------------------------------------------------------------

\begin{figure}[!h]
\begin{minipage}{0.3\textwidth}
 \centerline{\includegraphics[width=1.0\textwidth]{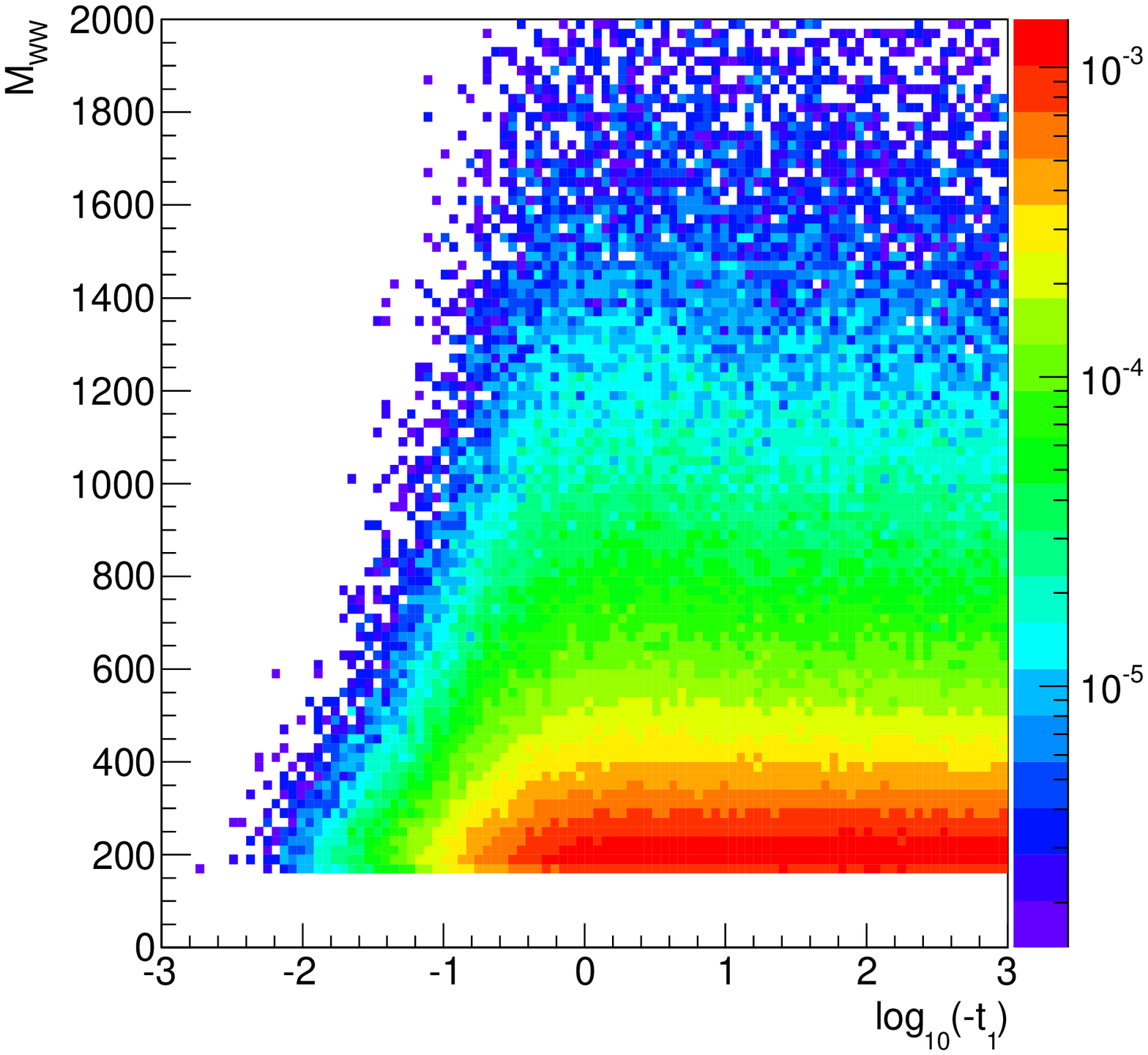}}
\end{minipage}
\hspace{0.2cm}
\begin{minipage}{0.3\textwidth}
 \centerline{\includegraphics[width=1.0\textwidth]{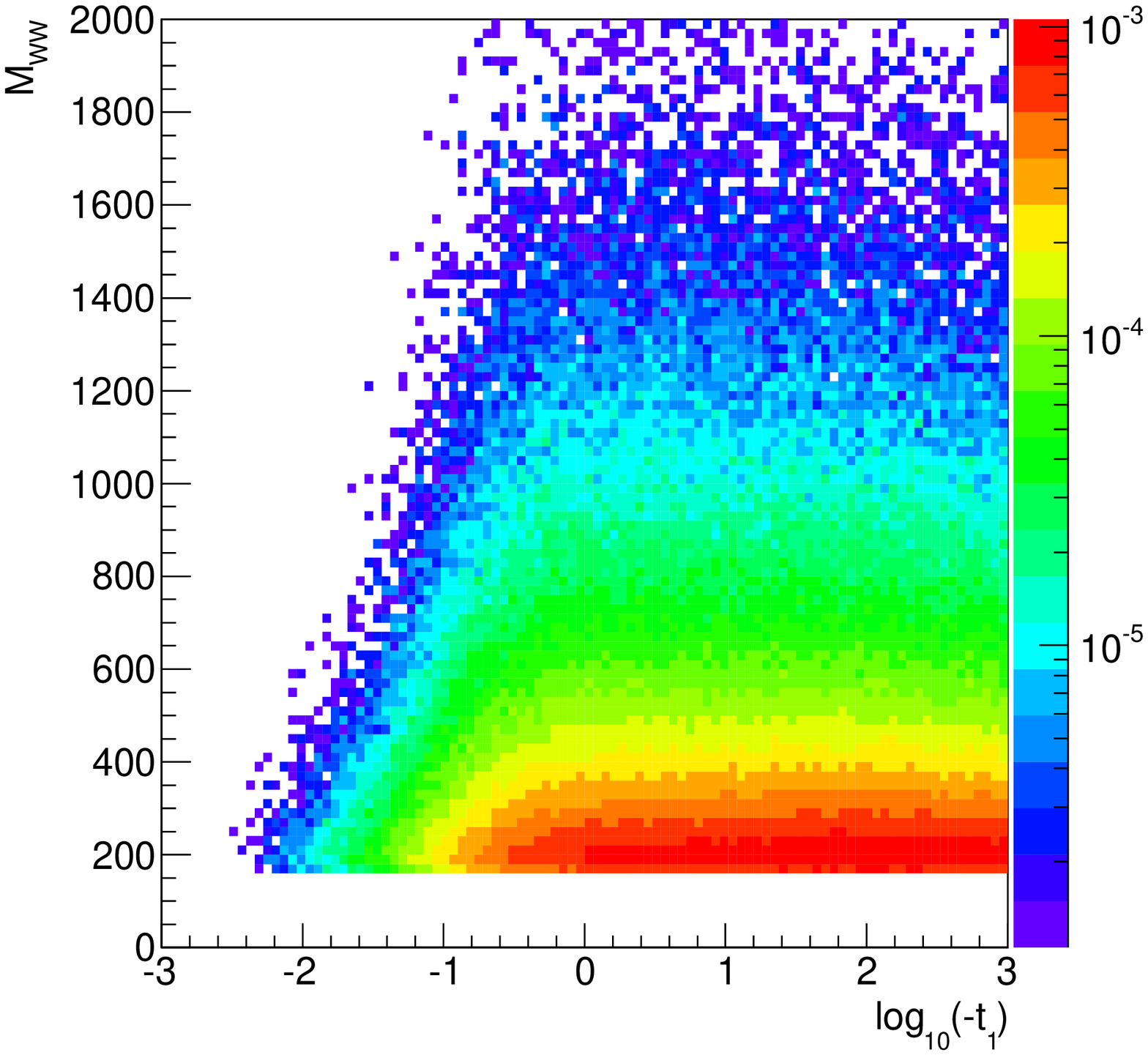}}
\end{minipage}
\hspace{0.2cm}
\begin{minipage}{0.3\textwidth}
 \centerline{\includegraphics[width=1.0\textwidth]{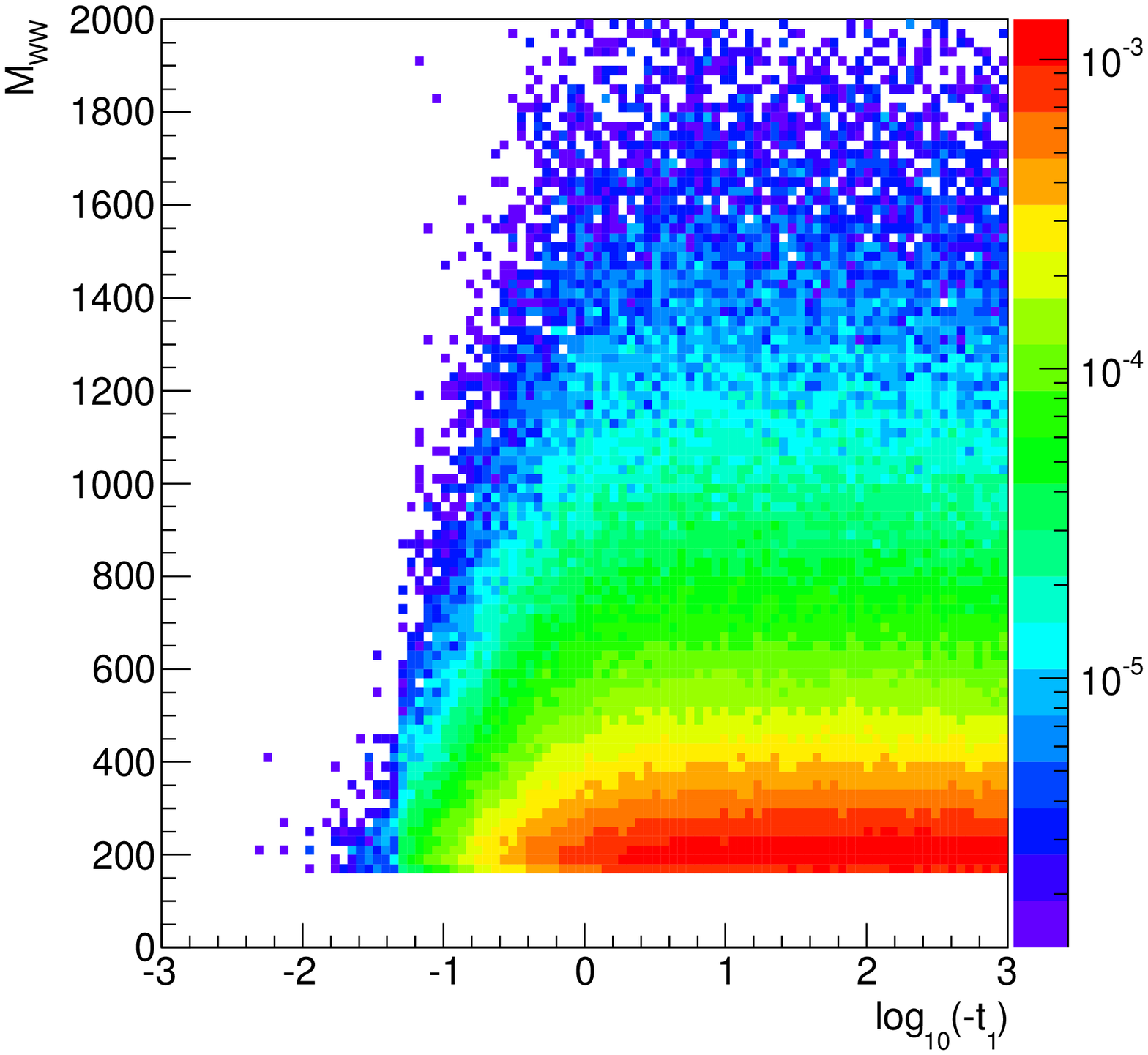}}
\end{minipage}

   \caption{Distributions in $Q_1^2 \times M_{WW}$ 
(or $Q_2^2 \times M_{WW}$) for different structure functions: 
LUX-like, ALLM97, SU for $\sqrt{s}$ = 13 TeV.
}
 \label{fig:Q2_MWW}
\end{figure}

In Fig.\ref{fig:Q2_MWW} we discuss correlation between $t_1 = -Q_1^2$
or $t_2 = -Q_2^2$ and invariant mass of the $W^+ W^-$ system produced
in the photon-photon fusion (please note logarithmic scale in rapidity).
At large $M_{WW}$ there are no small virtualities of photons. 
Therefore the collinear-factorization approach
may be expected to be better close to the threshold and worse for 
large $WW$ invariant masses.
This may be important in establishing a reference Standard Model result
in the studies searching for effects beyond Standard Model.
The result does not depend on the parametrization of the structure function.

%-----------------------------------------------------------------------%
%----------------------------------------------------------------------------

In the inelastic-inelastic case both protons undergo dissociation into
a complicated final state. What happens to the remnant systems will be 
discussed elsewhere. Here we show whether the photon virtualities and
Bjorken-$x$ values (arguments of the structure functions) are correlated.
Only a small correlation can be observed. The figure shows that rather
large Bjorken-$x$ give the dominant contribution. This is region
corresponding to fixed-target experiments performed in 80ies and 90ies.

%-----------------------------------------------------------------------%
\begin{figure}[!h]
\begin{minipage}{0.3\textwidth}
 \centerline{\includegraphics[width=1.0\textwidth]{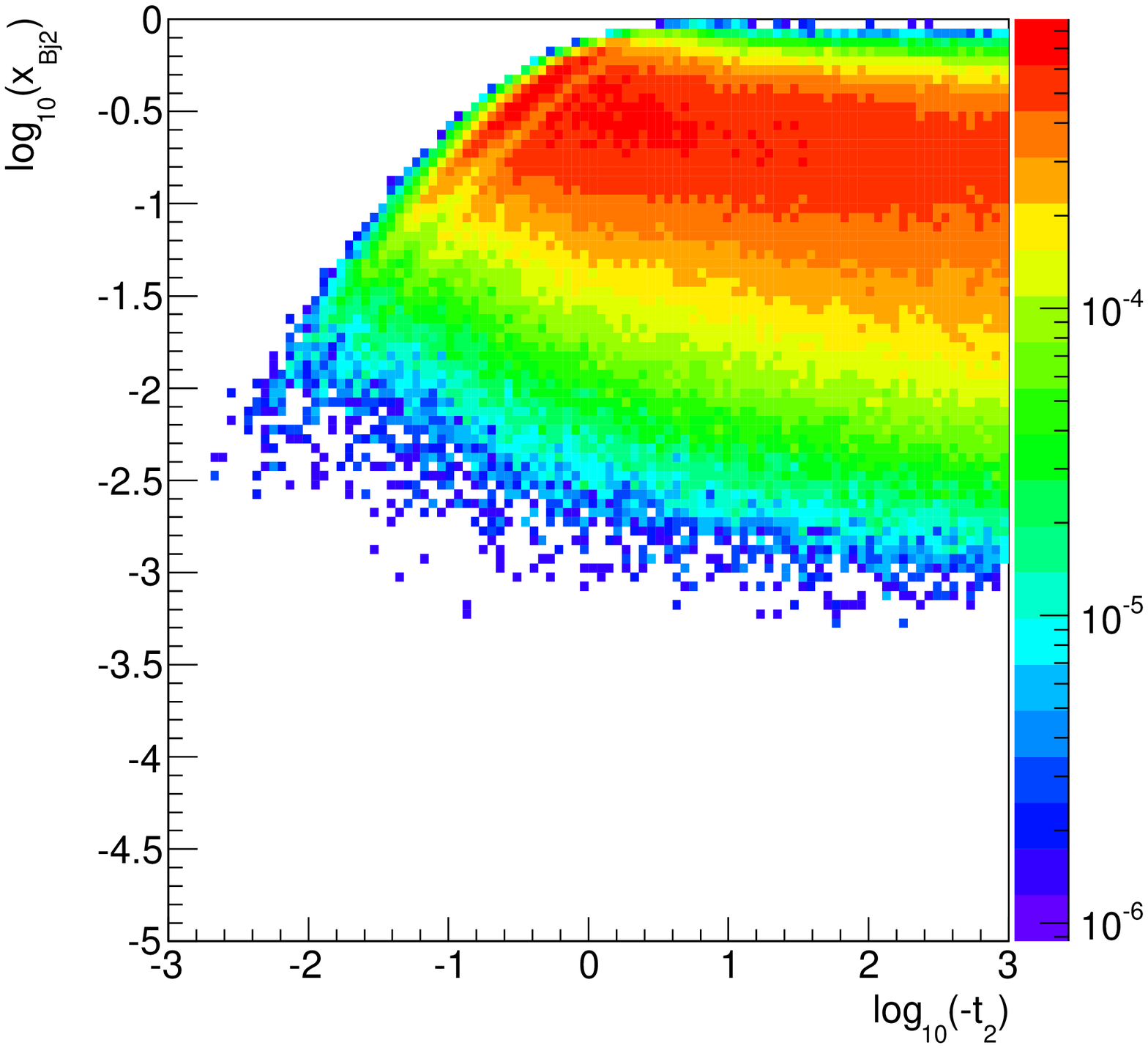}}
\end{minipage}
\hspace{0.2cm}
\begin{minipage}{0.3\textwidth}
 \centerline{\includegraphics[width=1.0\textwidth]{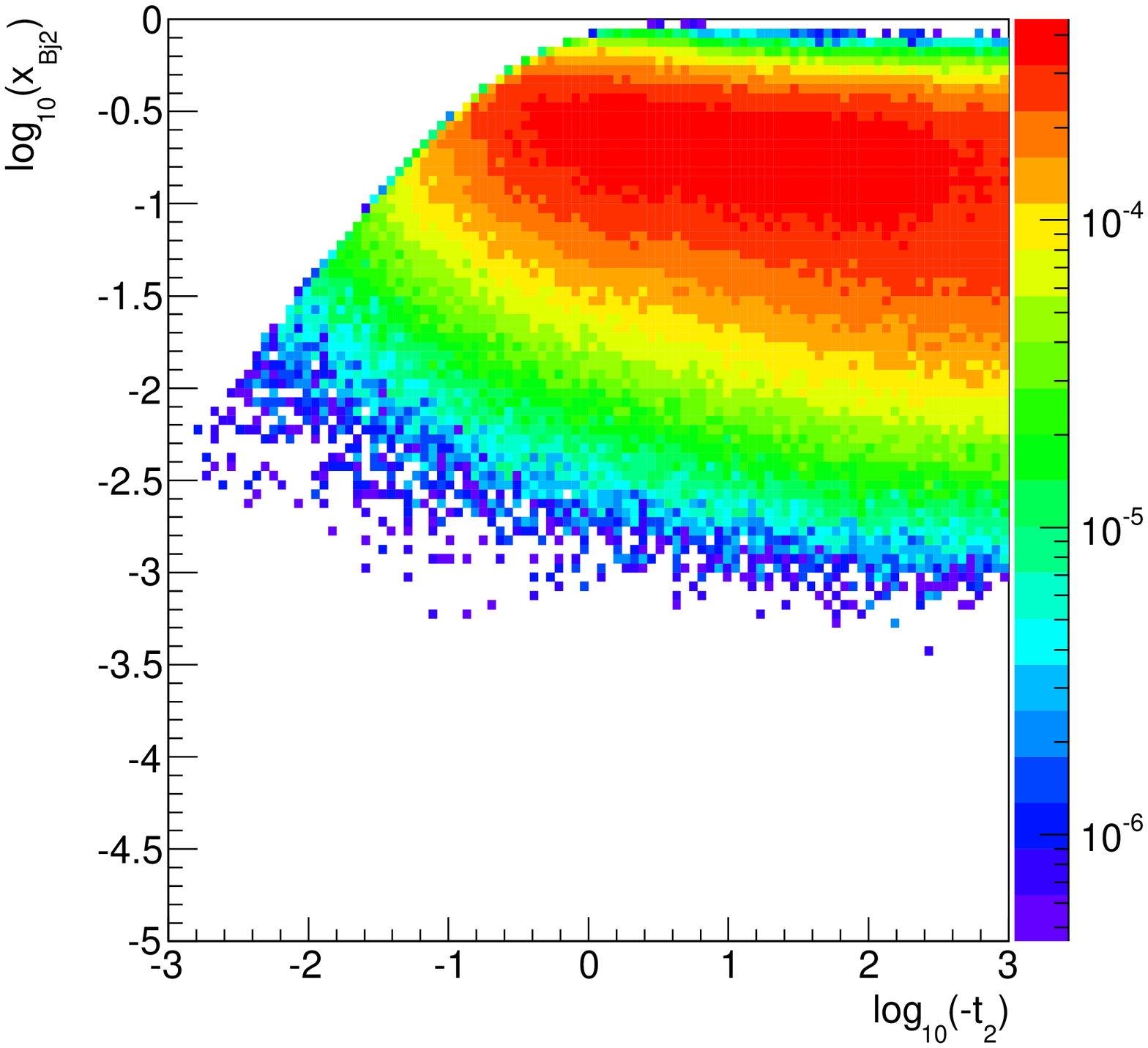}}
\end{minipage}
\hspace{0.2cm}
\begin{minipage}{0.3\textwidth}
 \centerline{\includegraphics[width=1.0\textwidth]{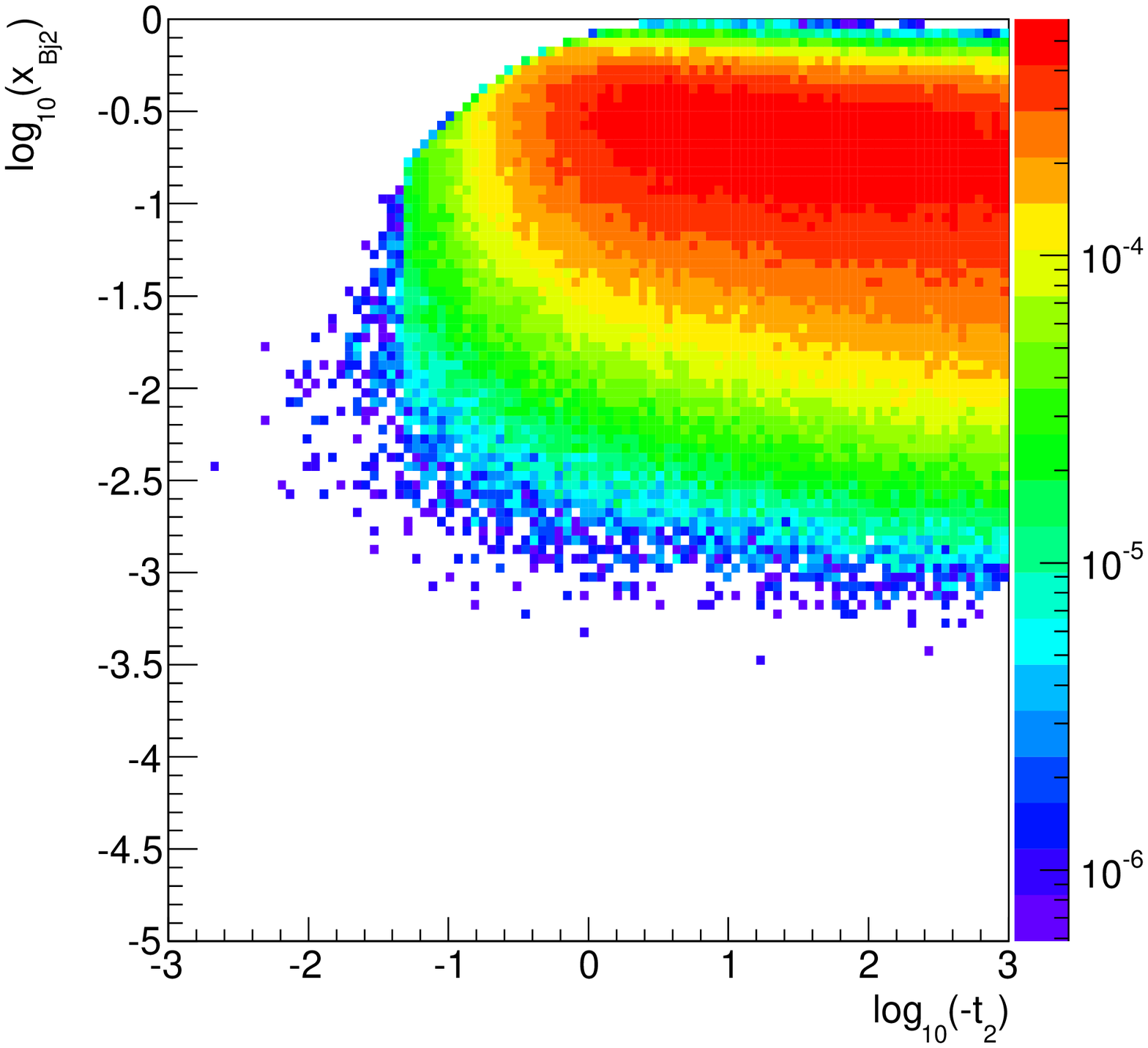}}
\end{minipage}

\caption{Correlations in $Q_{1/2}^2 \times x_{1/2}$ for different 
$F_2$ structure functions: LUX-like, ALLM97, SU for $\sqrt{s}$ = 13 TeV.
}
 \label{fig:dsig_Q2_x}
\end{figure}
%----------------------------------------------------------------------------

\begin{figure}[!h]
\begin{minipage}{0.3\textwidth}
 \centerline{\includegraphics[width=1.0\textwidth]{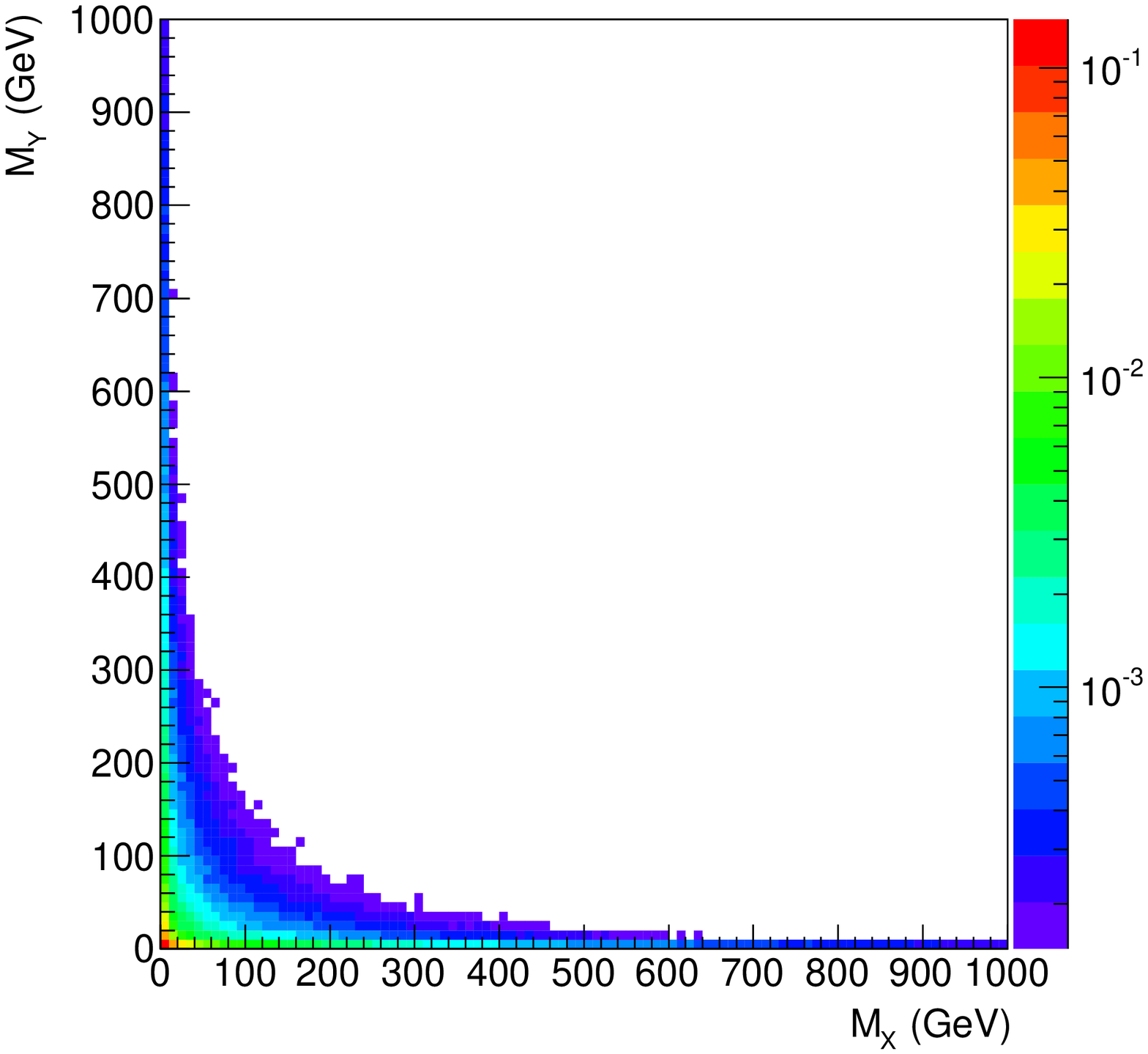}}
\end{minipage}
\hspace{0.2cm}
\begin{minipage}{0.3\textwidth}
 \centerline{\includegraphics[width=1.0\textwidth]{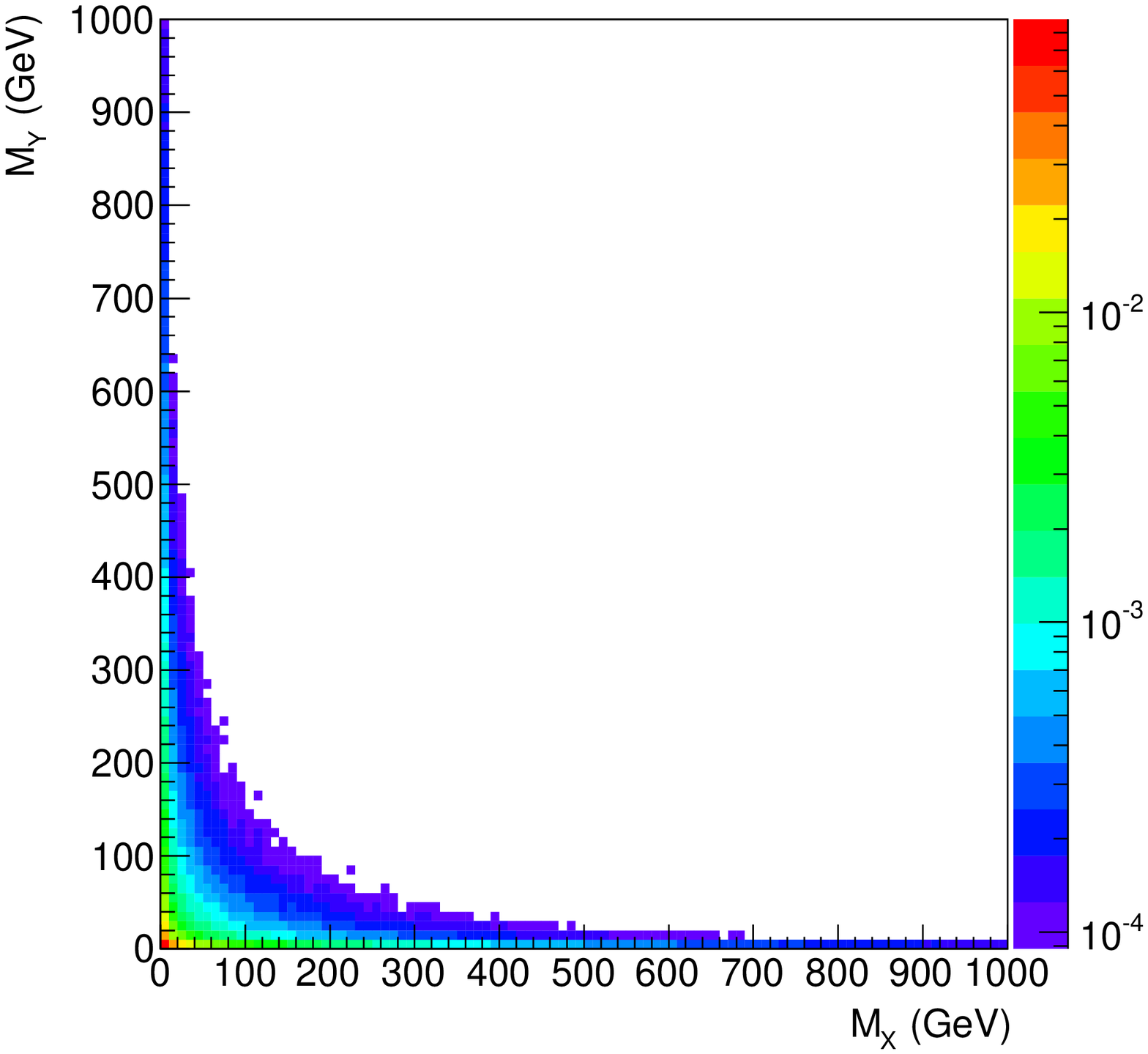}}
\end{minipage}
\hspace{0.2cm}
\begin{minipage}{0.3\textwidth}
 \centerline{\includegraphics[width=1.0\textwidth]{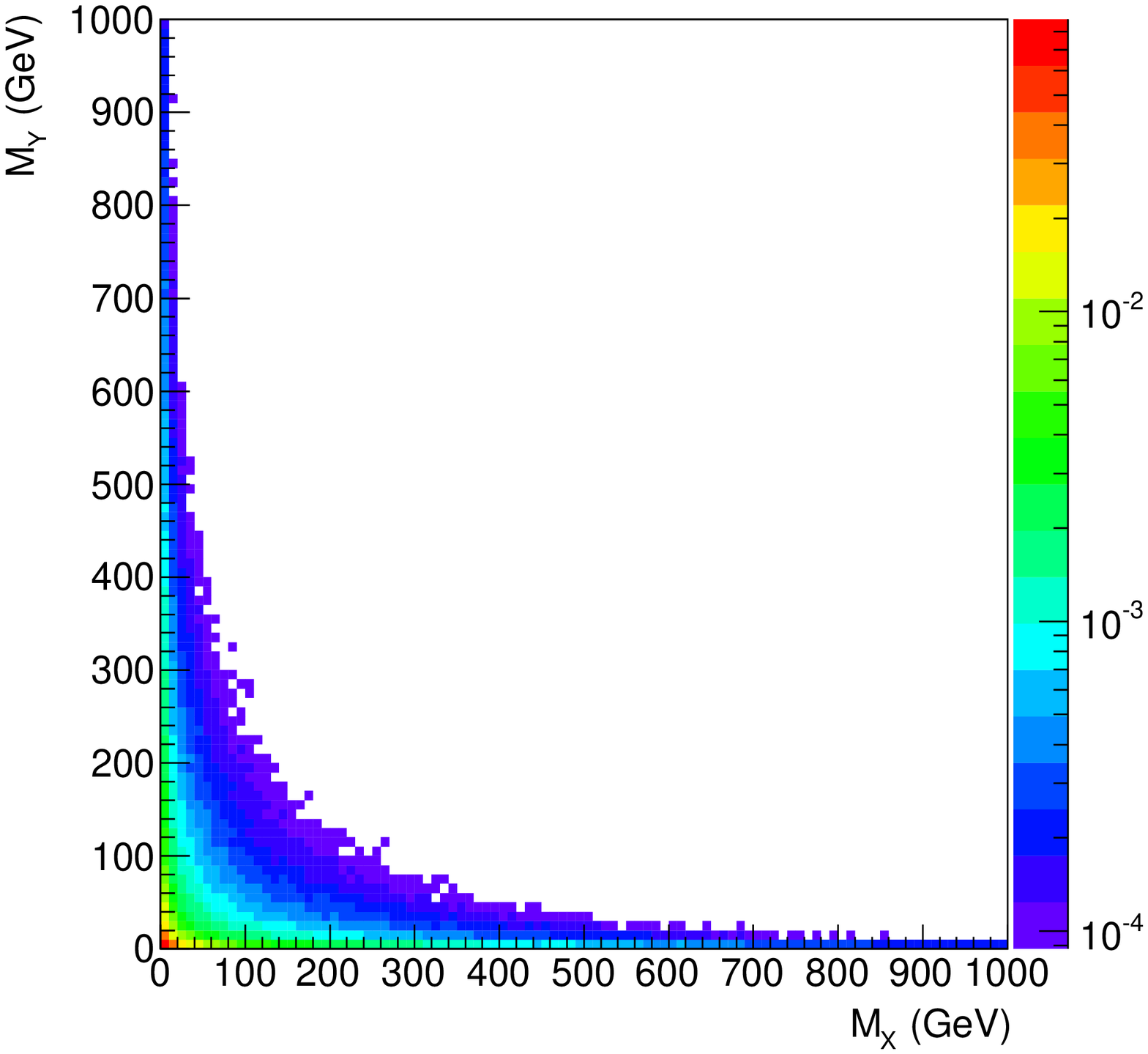}}
\end{minipage}

   \caption{Correlations in masses of the remnant systems 
for the double inelastic contribution for three different 
$F_2$ structure functions: LUX-like, ALLM97, SU for $\sqrt{s}$ = 13 TeV.
}
 \label{fig:dsig_dMXdMY}
\end{figure}

For completeness in Fig.\ref{fig:dsig_dMXdMY} we show potential
correlations in masses of both dissociated systems. The maximum
of the two-dimensional distribution occurs when $M_X, M_Y$
are rather small. When one of the masses is large the second 
is typically small.
So we typically expect situations with small rapidity gap on one side
and large gap on the other side of the ``centrally'' produced $W^+ W^-$
system. This will be discussed in detail elsewhere.

%-----------------------------------------------------------------------%
%----------------------------------------------------------------------------

%----------------------------------------------------------------------------
\subsection{Decomposition into polarization components}
%----------------------------------------------------------------------------

The matrix elements in Eq.(\ref{helicity_ME})
allow to calculate cross sections
for different states of polarization of $W$ bosons
(polarizations here are defined in the $W^+W^-$ center-of-mass frame, for explicit formulas,
see \cite{Nachtmann:2005en}).
It can be seen that the TT (both $W$'s are transversely polarized) component is
larger than 80 \%. The LL (both $W$'s longitudinally polarized) component
plays a special role in studies of $W W$ interactions.
However in the photon-photon fusion the cross section for
production of this component is smaller than 5 \% of the total cross section.

To make a thorough study of possible effects beyond the SM 
in the LL channel, one should include decays of $W$'s. 
Then, the small LL component can be enhanced by interference with
transverse $W$'s.

 %------------------------------------------------------------------------------------
%\begin{table}
\begin{table}[tbp]
\centering
\begin{tabular}{|c|c|c|c|}
\hline
contribution               &   8 TeV   &  13 TeV        \\
\hline        
TT                         &   0.405   &  0.950         \\
\hline
LL                         &   0.017   &  0.046         \\
\hline
LT + TL                & 0.028 + 0.028 & 0.052 + 0.052  \\            
\hline
SUM                        & 0.478 &  1.090   \\  
\hline
 
\end{tabular}
\caption{Contributions of different polarizations of $W$ bosons
for the inelastic-inelastic component for the LUX-like structure function. 
The cross sections are given in $p b$. 
}
\end{table}
%------------------------------------------------------------------------------------

\begin{figure}[!htbp]
\begin{minipage}{0.47\textwidth}
 \centerline{\includegraphics[width=1.0\textwidth]{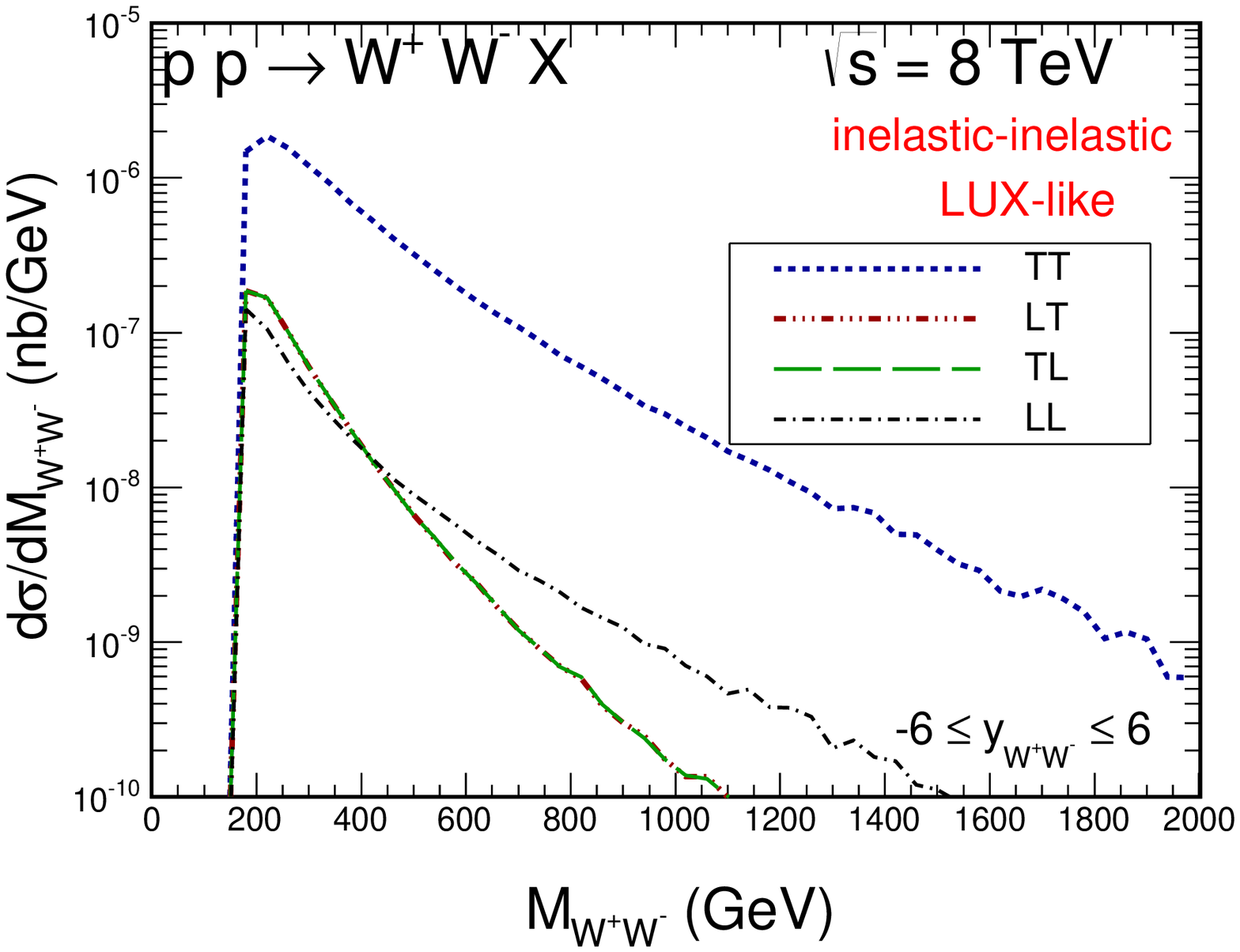}}
\end{minipage}
%\hspace{0.5cm}
\begin{minipage}{0.47\textwidth}
 \centerline{\includegraphics[width=1.0\textwidth]{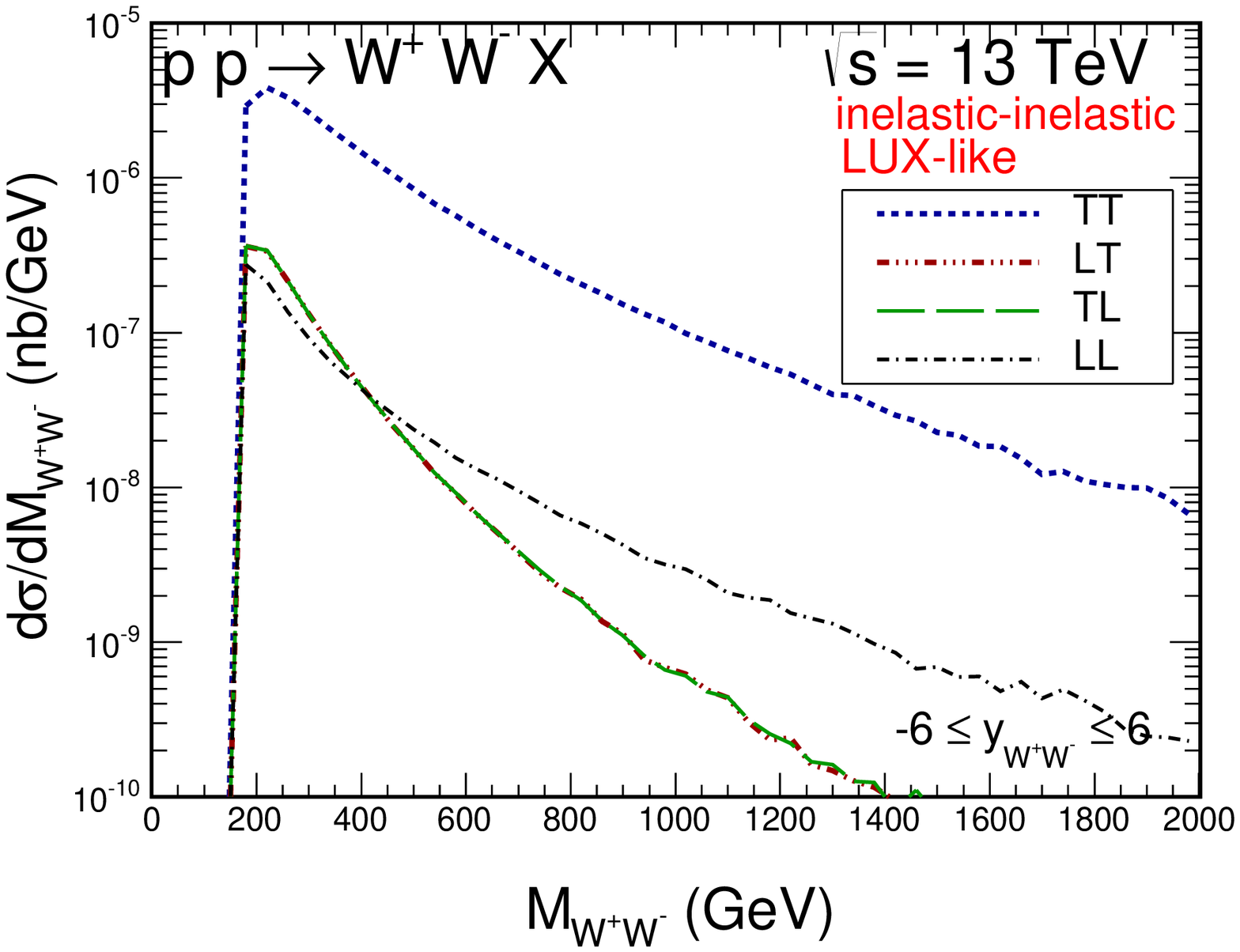}}
\end{minipage}
\caption{
\small
Decomposition into polarization states of W bosons for the inelastic-inelastic
component as a function of $M_{WW}$.
The calculation was performed for the LUX-like structure function. 
The left panel shows results for W = 8 TeV, while the right panel 
shows results for W = 13 TeV.
}
 \label{fig:dsig_dM_LT_decomposition}
\end{figure}

In fact it is more interesting what happens at large $W W$ invariant
masses $M_{WW} >$ 1 TeV where effect beyond Standard Model could show up.
In Fig.\ref{fig:dsig_dM_LT_decomposition} we show the decomposition
into different polarization states of W bosons as a function of the $W W$ invariant
mass. We observe that the $TT$ component dominates in the whole
invariant mass region.

%-----------------------------------------------------------------------------
%------------------------------------------------------------------------------

%----------------------------------------------------------------------
\subsection{Role of longitudinal structure function}
%----------------------------------------------------------------------

We now wish to discuss the importance of the longitudinal structure
function in the photon-flux. 
This needs some clarification. Arguably the most physical representation
of the inelastic flux  would be 
to write the inelastic flux \ref{eq:flux_in} directly in terms of 
structure functions $F_T(x_{\rm Bj},Q^2) = 2 x_{\rm Bj} F_1(x_{\rm Bj},Q^2)$
and $F_L(x_{\rm Bj},Q^2)$. In terms of these structure functions 
$F_2$ decomposes
as $F_2(x_{\rm Bj},Q^2) =(F_T(x_{\rm Bj},Q^2) + F_L(x_{\rm Bj},Q^2)  )/(1 + \kappa^2)$,
with $\kappa^2 = 4 x_{\rm Bj}^2 m_p^2/Q^2$.
If we insert this into Eq.(\ref{eq:flux_in}), we get positive
contributions from $F_T$ as well as $F_L$.
In practice, we have a wealth of experimental data on $F_2$, and much less
knowledge of $F_L$. It is therefore more practical to express the photon 
flux directly in terms of $F_2$ and $F_L$.

\begin{figure}[!htbp]
\begin{minipage}{0.47\textwidth}
 \centerline{\includegraphics[width=1.0\textwidth]{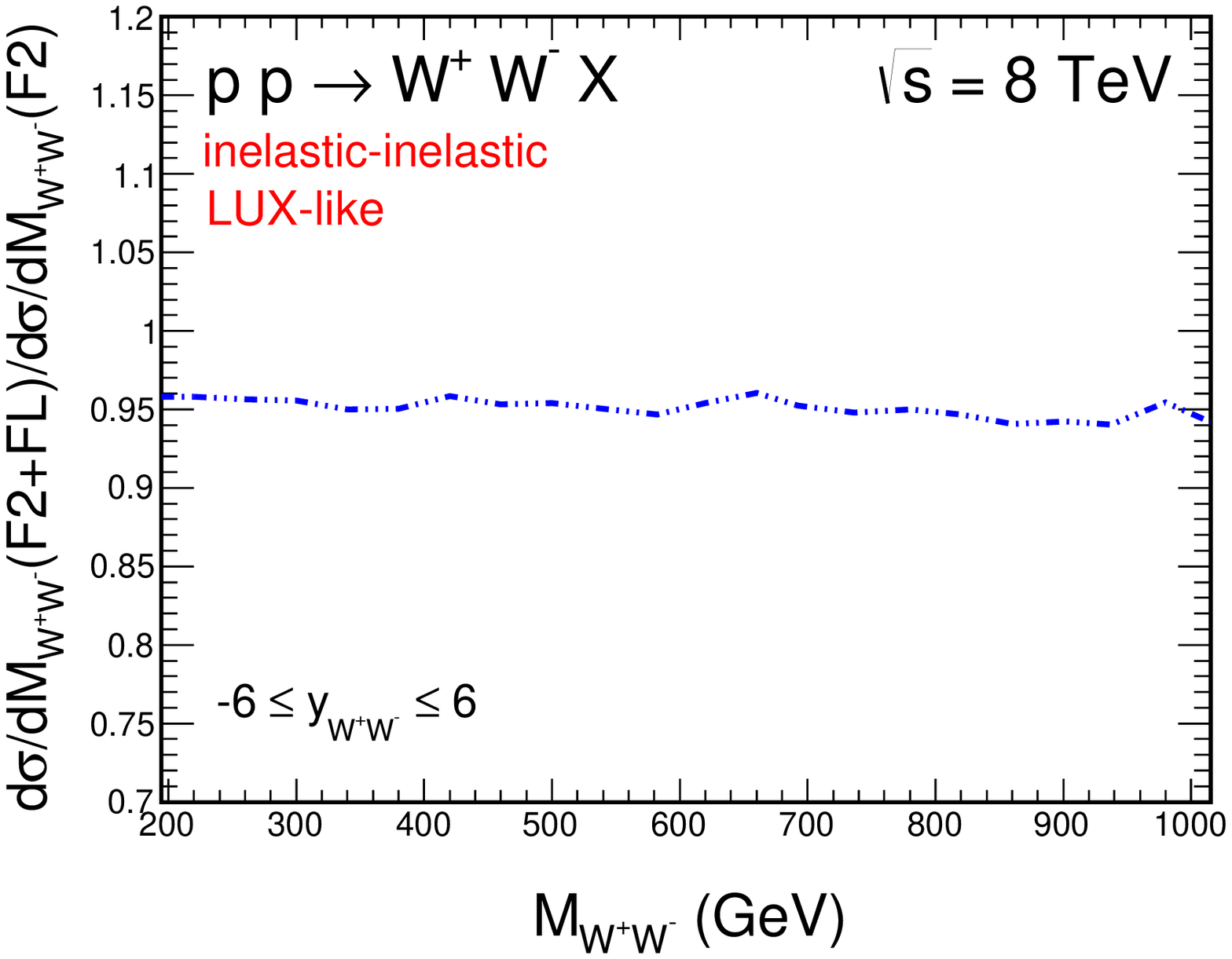}}
\end{minipage}
%\hspace{0.5cm}
\begin{minipage}{0.47\textwidth}
 \centerline{\includegraphics[width=1.0\textwidth]{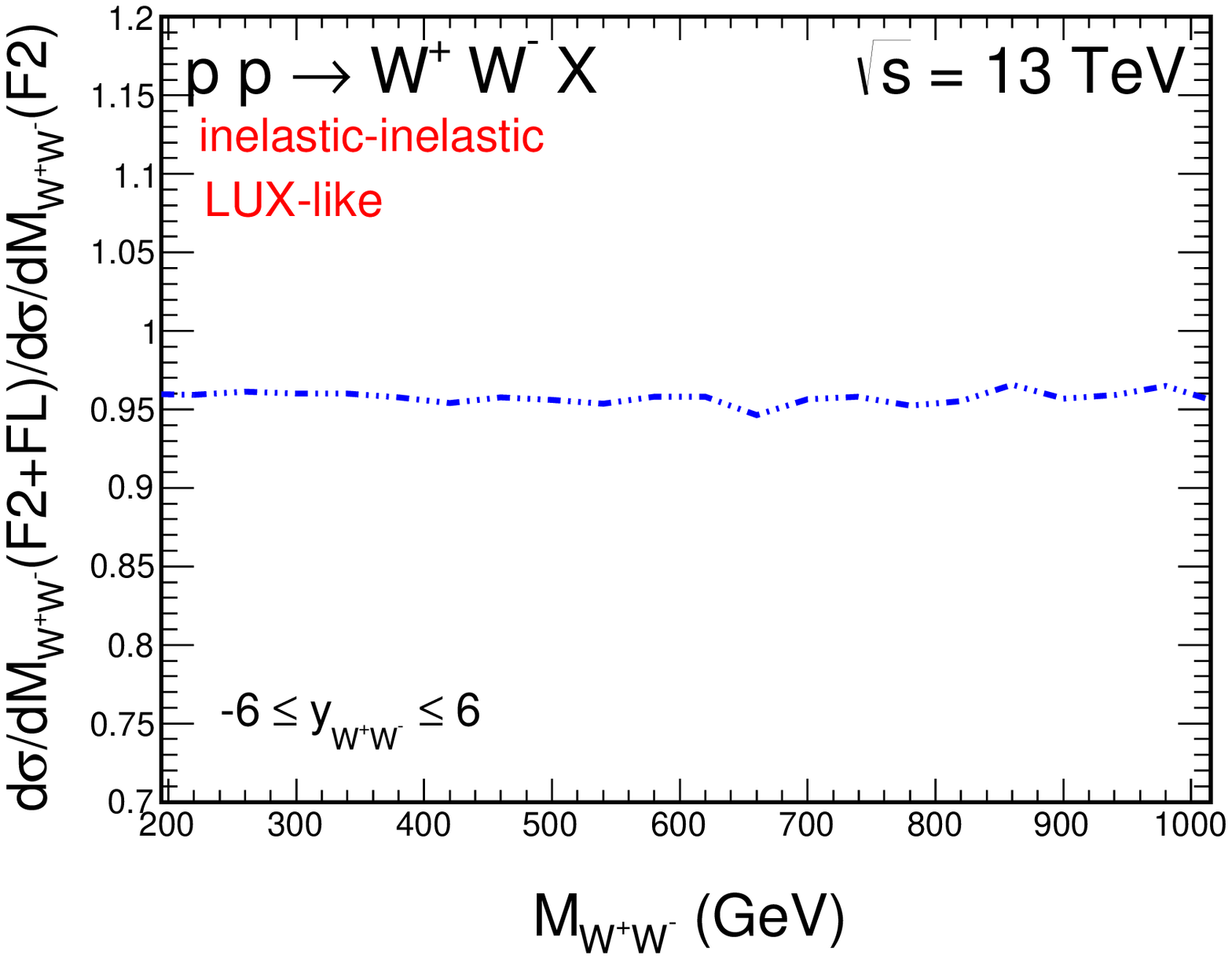}}
\end{minipage}
\caption{
\small
The role of the longitudinal structure function as a function of
$W^+ W^-$ invariant mass.
Shown is the ratio of the cross section with and without $F_L$ structure
function in the unintegrated photon fluxes.
The calculation was performed for the LUX-like structure function. 
The left panel shows results for W = 8 TeV, while the right panel 
shows results for W = 13 TeV.
}
 \label{fig:dsig_dM_FL}
\end{figure}

We now want to check to which extent the photon fluxes can be evaluated 
from $F_2$ only.
We therefore evaluate the photon flux for two different cases:
\begin{enumerate}
	\item 
	in Eq.(\ref{eq:flux_in}) we substitute 
	$2 x_{\rm Bj} F_1(x_{\rm Bj},Q^2) = ( 1+ \kappa^2 ) F_2(x_{\rm Bj},Q^2) - F_L(x_{\rm Bj},Q^2)$ 
	(denoted as $d\sigma(F_2 + F_L)/dM_{WW}$),
	\item 	in eq.(\ref{eq:flux_in}) we substitute $2 x_{\rm Bj} F_1(x_{\rm Bj},Q^2) = F_2(x_{\rm Bj},Q^2)$ (denoted as  $d\sigma(F_2)/dM_{WW})$.
\end{enumerate}
In Fig.\ref{fig:dsig_dM_FL} we show the ratio 
$d\sigma(F_2 + F_L)/dM_{WW}/d\sigma(F_2)/dM_{WW}$ for two different energies.
In such a decomposition the cross section when both $F_2$ and $F_L$
are taken into account is smaller by 4-5 \% than the cross section when 
only $F_2$ is taken into account, independent of $M_{WW}$. 

%-----------------------------------------------------------------------------
%------------------------------------------------------------------------------

%---------------------------------------------------------------
\subsection{Rapidity distance between $W$ bosons}
%---------------------------------------------------------------

The $\gamma \gamma$ contribution is of the order of 2 \% for the inclusive
cross section as discussed at the beginning of this section. 
The technical problem is how to measure the $\gamma \gamma$ contribution 
in experiment. 
This can be done by imposing an extra condition on the size of the 
rapidity gaps around the electroweak vertex.

\begin{figure}[!htbp]
\begin{minipage}{0.47\textwidth}
 \centerline{\includegraphics[width=1.0\textwidth]{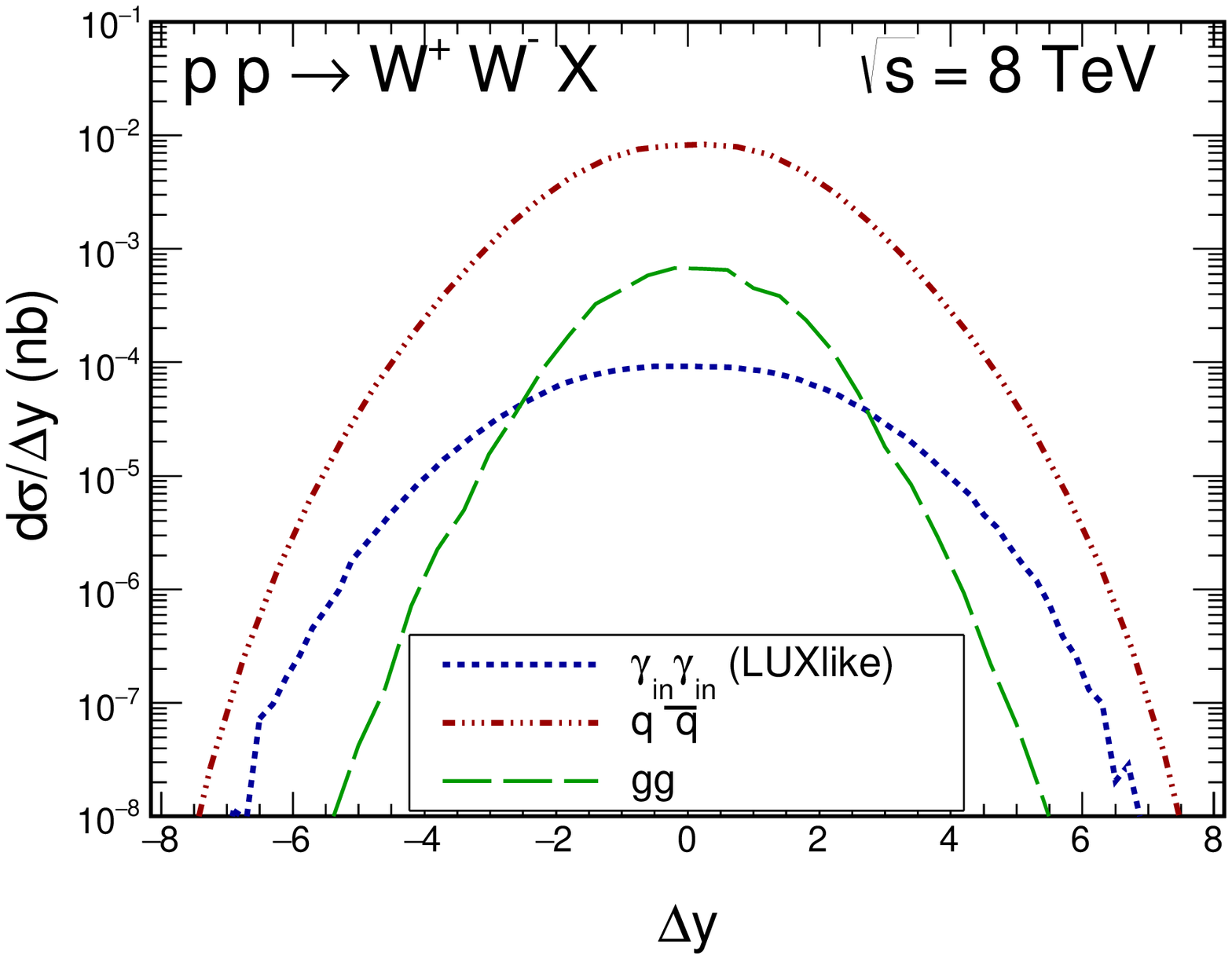}}
\end{minipage}
%\hspace{0.5cm}
\begin{minipage}{0.47\textwidth}
 \centerline{\includegraphics[width=1.0\textwidth]{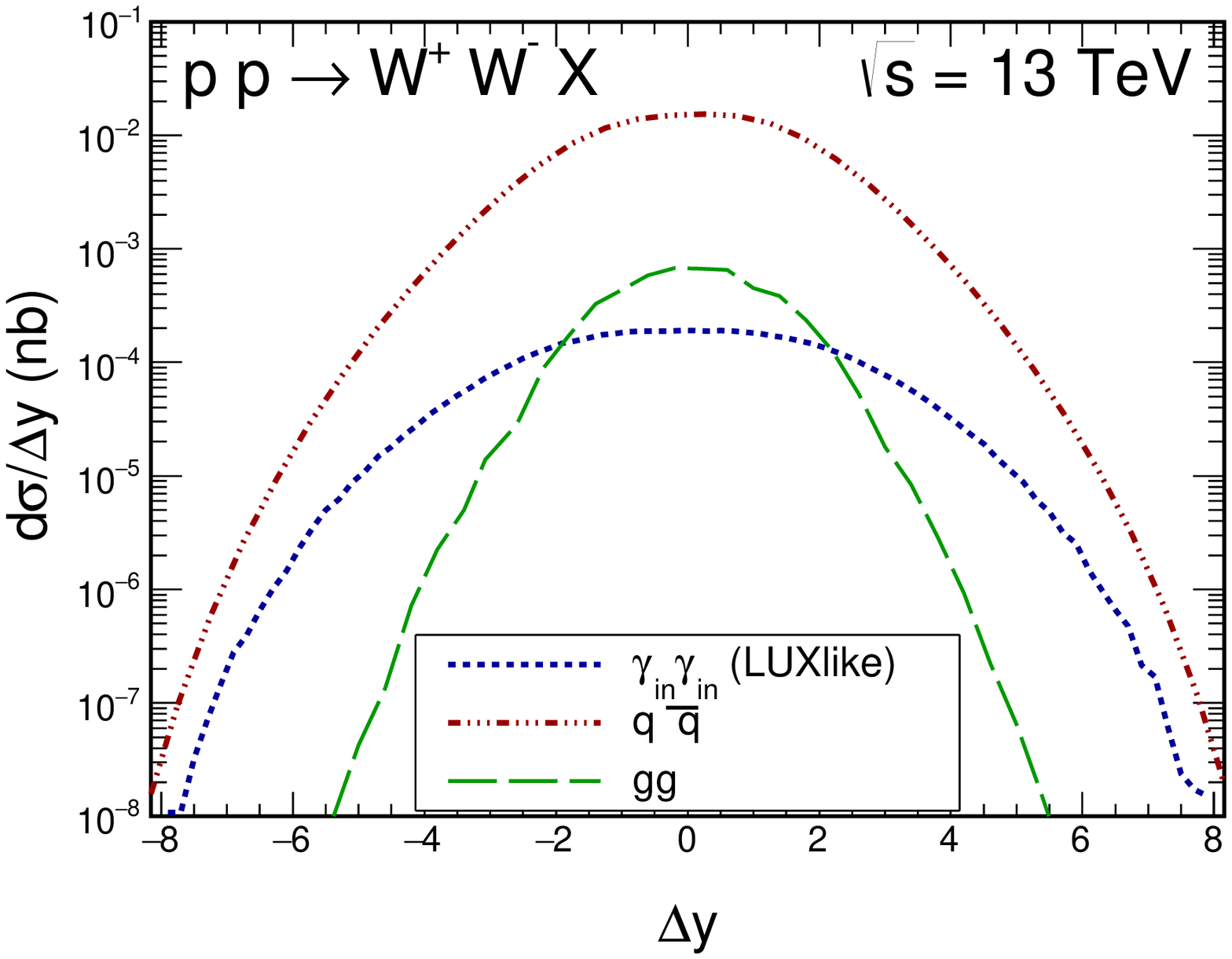}}
\end{minipage}
\begin{minipage}{0.47\textwidth}
 \centerline{\includegraphics[width=1.0\textwidth]{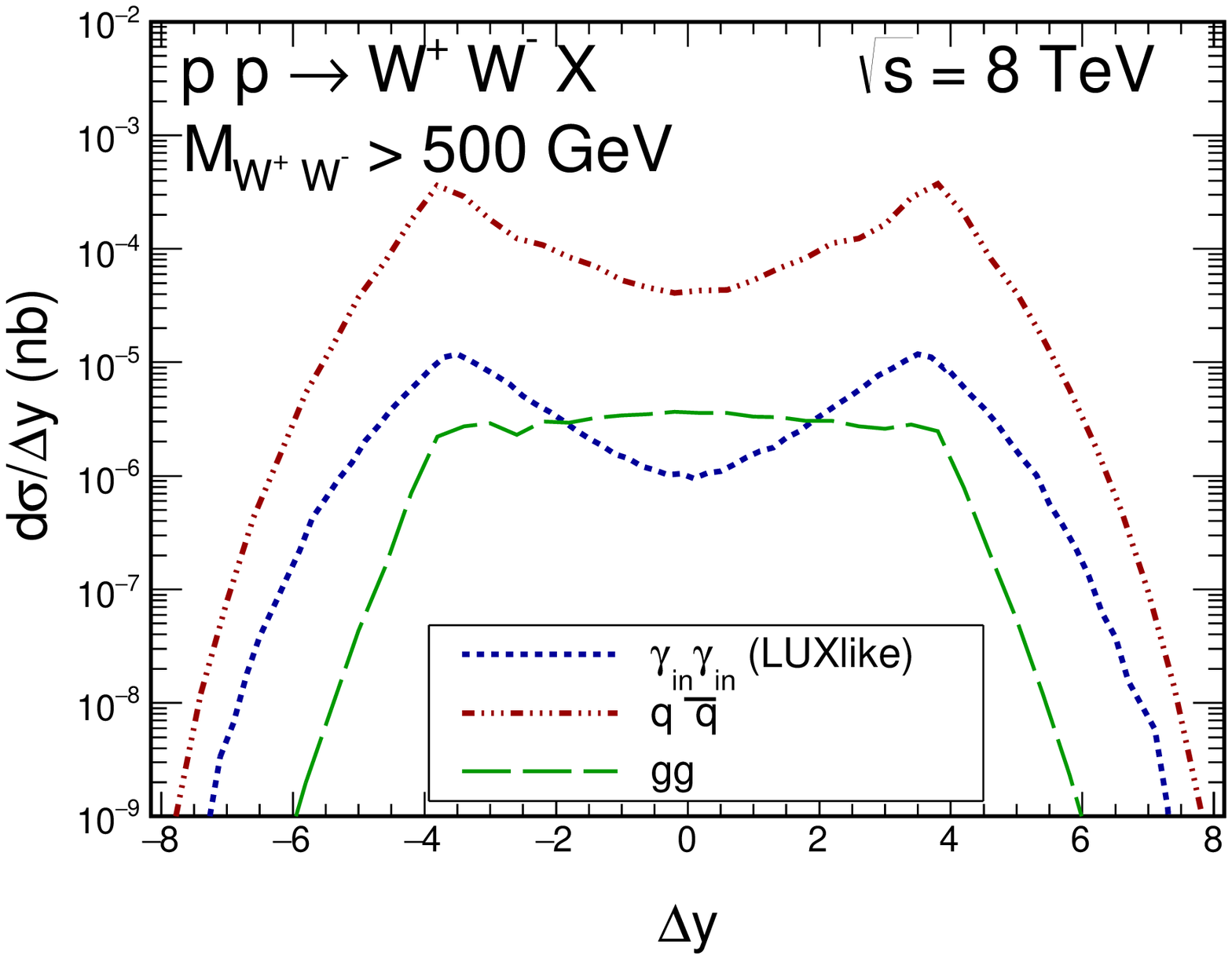}}
\end{minipage}
%\hspace{0.5cm}
\begin{minipage}{0.47\textwidth}
 \centerline{\includegraphics[width=1.0\textwidth]{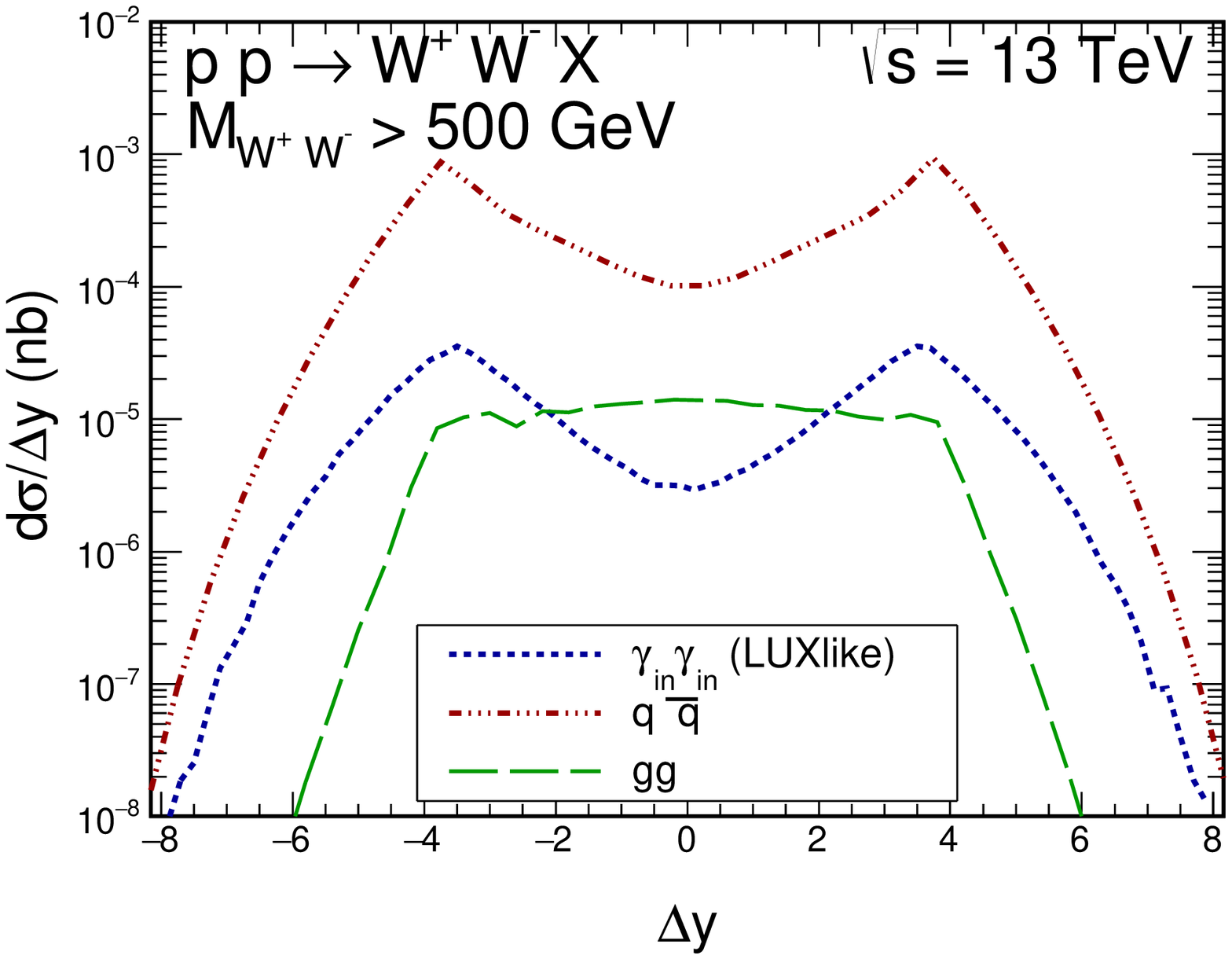}}
\end{minipage}
\caption{
\small
Distribution in rapidity distance between $W$ bosons.
The calculation for the $\gamma-\gamma$ contribution
(dotted-line, inelastic-inelastic contribution only) was performed 
for the LUX-like structure function. 
The left panel shows results for W = 8 TeV, while the right panel 
shows results for W = 13 TeV. For comparison we show also contribution
of the $q \bar q$, $\bar q q$ annihilation (dash-dotted line) and 
$g g \to W^+ W^-$ (dashed line). In the lower panels we show results for
extra cut imposed on the invariant mass of the $M_{WW}$ system --
$M_{WW} >$ 500 GeV.
}
 \label{fig:dsig_dydiff}
\end{figure}

In Fig.\ref{fig:dsig_dydiff} we show the distribution in the distance
in rapidity between the two produced $W$ bosons (dotted line)
without any extra condition on rapidity gaps. 
The distribution is fairly flat over several units.
This (rapidity distance between muon and electron) can perhaps 
be used to enhance the data sample for the
$\gamma \gamma \to W^+ W^-$ mechanism. 
For reference we show also contribution of the $q \bar q + \bar q q$ 
annihilation (dash-dotted line) and gluon-gluon fusion (dashed line) which 
proceed via quark loops. The latter calculation is performed with LoopTools
package \cite{Hahn:1998yk} (for details we refer to \cite{LS2013}).
The distribution corresponding to gluon-gluon fusion is much narrower
than that of the $\gamma \gamma$ fusion. It is not so for 
the quark-antiquark annihilation. The latter is broader due to
parton distribution product ($q(x_1) \bar q(x_2)$, containing valence quarks) 
as well as due to presence of $s$-channel photon and $Z$-boson exchanges.
Excluding artificially the latter contributions makes the distribution
in $\Delta y$ much narrower.
The distributions change shapes when imposing extra cut on $M_{WW} >$
500 GeV (lower panels), but the general situation is similar.

%-----------------------------------------------------------------------------
%------------------------------------------------------------------------------

%--------------------------
\section{Conclusions}
%--------------------------

In the present paper we have discussed the production of $W^+ W^-$ pairs
created via the photon-photon fusion mechanism. 
In contrast to previous approaches we include transverse momenta 
of photons incoming to the hard process.
The matrix elements derived in \cite{Nachtmann:2005en} have been used.
The explicit dependence on polarization state of $W$ bosons has allowed us
to calculate different polarization contributions.

We have obtained cross section of about 1 pb for the LHC energies.
This is about 2 \% of the total integrated cross section dominated
by the quark-antiquark annihilation and gluon-gluon fusion.

Different combinations of the final states (elastic-elastic,
elastic-inelastic, inelastic-elastic, inelastic-inelastic) related to
whether the incoming protons do or do not undergo dissociation
have been considered. We have focused rather on the dominant 
inelastic-inelastic component.

The unintegrated photon fluxes were calculated based on modern
parametrizations of the proton structure functions from the literature.

Several differential distributions in $W$ boson transverse momentum
and rapidity, $WW$ invariant mass, transverse momentum of the $WW$ pair
have been presented and compared with previous results obtained in
the collinear approach in \cite{LSR2015}.
We have obtained a smaller cross section for large $W^+ W^-$ invariant
masses than in the collinear approximation. Our predictions may be
considered as realistic Standard Model reference in searches
of effects beyond Standard Model in the $\gamma \gamma \to W^+ W^-$
process.

Several correlation observables have been studied. 
Large contributions from the regions of large photon virtualities 
$Q_1^2$ and/or $Q_2^2$ have been found putting in question the
reliability of leading-order collinear-factorization approach.
We have found larger virtualities for larger invariant masses
of the $W^+ W^-$ system. This results seems universal and would be
similar e.g. for production of charged Higgs $H^+ H^-$ pairs via $\gamma \gamma$
fusion.

We have found that $x$ values (arguments of $F_2$ structure functions)
are typically $x \sim$ 0.1-0.5. 
In contrast to the production of charged lepton pairs the production 
of $W^+ W^-$ pairs requires therefore structure functions
in the region where they were studied (measured and fitted).
The dominant part comes from the region described by the DGLAP
evolution equation and only a small fraction comes from nonperturbative
region. The nonperturbative contribution (small $Q^2$ region) was
much larger for the charged lepton production
\cite{Luszczak:2015aoa} where a detailed studies 
of resonances was necessary.

We have presented a decomposition of the cross section into
individual contributions of different polarizations of both $W$ bosons.
It has been shown that the $TT$ (both $W$ transversally polarized)
contribution dominates and constitutes
a little bit more than 80 \% of the total cross section.
The $LL$ (both $W$ longitudinally polarized) contribution 
is interesting in the context of studying
$W W$ interactions or searches beyond the Standard Model.
However, the corresponding cross section is only about 5 \%. 
We have found only a mild dependence of relative amount of different
contributions as a function of $WW$ invariant mass.

We have quantifield the effect of inclusion of longitiudinal structure function 
into the transverse momentum dependent fluxes of photons. A rather small, approximataly
$M_{WW}$~-~independent, effect was found.

The discussed here $\gamma \gamma \to W^+ W^-$ mechanism leads to rather large
rapidity separations of $W^+$ and $W^-$ boson. It requires further studies to understand
whether it can be used 
to relatively enhance contribution of the $\gamma \gamma \to W^+ W^-$ in experimental
studies.
\vspace{1cm}

{\bf Acknowledgments}

We are indebted to Piotr Lebiedowicz for providing us a program
to calculate gluon-gluon fussion mechanism.
This study was partially supported by the Polish National Science Centre 
grants DEC-2013/09/D/ST2/03724 and
DEC-2014/15/B/ST2/02528 and by the Center for Innovation and
Transfer of Natural Sciences and Engineering Knowledge in Rzesz{\'o}w.
We are indebted to Laurent Forthomme for discussion of
some issues presented here.

%-------------------------------------------------------------------------------------

\end{document}